\documentclass{aastex631}

\newcommand{\detectionthreshold}{5}

\newcommand{\boxcarwindowsize}{$500\,{\rm s}$}
\newcommand{\totalnumvortices}{815}
\newcommand{\numnewvortices}{506}

\newcommand{\discardedvortices}{65}

\newcommand{\numvorticeswithDeltaT}{112}
\newcommand{\DeltaTSNR}{3}

\received{2022 Mar 31}
\revised{2022 Jul 2}
\accepted{2022 Jul 19}

\submitjournal{PSJ}

\shorttitle{Estimating the Heights of Martian Vortices from Mars 2020 MEDA Data}
\shortauthors{Jackson}

\graphicspath{{./}{}}

\begin{document}

\title{Estimating the Heights of Martian Vortices from Mars 2020 MEDA Data}

\author[0000-0002-9495-9700]{Brian Jackson}
\affiliation{Department of Physics\\ 
Boise State University\\ 
1910 University Drive, Boise ID 83725-1570 USA}

\begin{abstract}
Small convective vortices occur ubiquitously on Mars, frequently as dust devils, and they produce detectable signals in meteorological data -- in pressure, temperature, and wind speed and direction. In addition to being important contributors to the martian dust budget, convective vortices may serve as probes of the boundary layer, providing clues on convective instability, boundary layer diurnal evolution, and surface-atmosphere interactions. Using vortices as boundary layer probes requires a detailed understanding of the link between their properties and occurrence rates and the conditions that produce them. Fortunately, the growing cache of data from the Mars Environmental Dynamics Analyzer (MEDA) instrument suite onboard the Mars 2020 Perseverance rover promises to elucidate these relationships. In this study, we present a catalog of vortex detections from mission sols 90 through 179 to bolster our previous catalog based on sols 15 through 89. Consistent with predictions, we find more vortex encounters during this second half of the mission than from the first half. In addition to analyzing the pressure signals from these vortex encounters, we also use a Gaussian process analysis to recover contemporaneous temperature signals. By combining these signals with a long-established thermodynamics model, we estimate heights of the vortices and find some agreement with previous work and evidence for the diurnal growth and decay of the martian boundary layer. We also discuss prospects for additional boundary layer studies using martian vortex encounters. 
\end{abstract}


\section{Introduction} \label{sec:Introduction}

Dust devils appear ubiquitously on Mars, lofting a significant fraction of the dust in the martian atmosphere \citep{2016SSRv..203...89F}. Several studies, many involving terrestrial analog field work, have explored the conditions that produce dust devils \citep{2016SSRv..203...39M}. \citet{1970JGR....75..531R} reported a month-long field campaign in the Mojave Desert in a broad basin continually groomed to provide plentiful dust for dust devil formation and found, for example, that their occurrence rate correlated with high near-surface lapse rates, presumably because such high lapse rates give rise to convective instability. Landed spacecraft have enabled similar field campaigns on Mars \citep[e.g.,][]{2010JGRE..115.0E16E}, albeit usually with much more limited instrumentation and data volume. Frequently including pressures, temperatures, wind speeds, etc., these rich martian datasets have highlighted some of the conditions under which dust devils frequently occur and under which they do not \citep{2021PSJ.....2..206J}.

Mounting evidence suggests that dust devil formation is highly sensitive to conditions within the planetary boundary layer (PBL). \citet{1990JApMe..29..498H} collated a list of conditions associated with dust devil formation, including the requirement that the PBL depth must exceed the Obokhuv length in magnitude by at least a factor of 100. (The Obokhuv length parameterizes the tendency for a fluid to convect and reaches negative values as convective instability sets in.) As the base of a planetary atmosphere warms from morning to afternoon, the PBL deepens \citep[cf.][]{2001imm..book.....A}, promoting dust devil formation. Such growth also promotes formation of small convection cells generally, whether they are visualized by lofted dust (dust devils) or not (dustless devils -- \citealp{2016Icar..278..180S}). Additional requirements include the availability of dust and sufficiently high vortex wind speeds.

The sensitivity of dust devil formation to ambient conditions has suggested their use as a probe of near-surface meteorology. \citet{2015Icar..260..246F}, for instance, conducted a comprehensive survey of space-based images of martian dust devils and found that the inter-devil spacing seems to match closely the PBL depth, opening the door to direct estimation of the PBL depth without landed or space-based meteorological instrumentation; orbital imagery of active dust devils might suffice. \citet{2015Icar..260..246F} also found a strong correlation between PBL depth and dust devil height, with the dust columns mounting to about one-fifth the PBL depth expected from meteorological modeling \citep{2013JGRE..118.1468C}. In fact, the correlation between dust devil height and PBL depth probably is probably linked to vortex thermodynamics.

In a seminal work, \citet{1998JAtS...55.3244R} devised a model of convective vortices as heat engines, with heating from a warm surface producing a temperature contrast $\Delta T_c$ at the vortex center relative to the surrounding environmental temperature $T_\infty$. The resulting buoyancy is assumed to drive convective motion and intake of air near the surface, resulting in a small pressure contrast $\Delta P_c$ compared to the ambient pressure $P_\infty$ and some frictional loss of energy throughout the system. (The fraction of energy lost near the surface as opposed to at altitude is parameterized as $\gamma \sim 1$.) The ascending column is assumed to reach a well-defined height $h$ at which point the warm air radiates into the surrounding environment before cooling and descending to complete the convective circuit. \citet{2020Icar..33813523J} expanded on that model to formulate a simple relationship between the vortex's height and radius, in the limit that the height was small compared to the atmospheric scale height $H$:

\begin{equation}
    \frac{h}{H} \approx \left( \frac{2\chi}{\gamma \left( 1 + \chi \right) } \right) \left( \frac{\Delta P_{\rm c}}{P_\infty} \right) \left( \frac{T_\infty}{\Delta T_{\rm c}}, \right)\label{eqn:vortex_height}
\end{equation}
where $\chi$ is the ratio of the specific gas constant to the specific heat at constant pressure (i.e., $\chi = R/c_{\rm P}$). (Equation 10 from \citealt{2020Icar..33813523J} on which the above equation is based mistakenly excluded the term $\chi/(1 + \chi)$ -- see Appendix \ref{sec:corrected_derivation} for a corrected derivation.) Uncertainties on the pressure and temperature parameters in Equation \ref{eqn:vortex_height} necessarily contribute to uncertainties on $h/H$. In particular, since $\Delta T_{\rm c}$ values are relatively small (few K), accurate assessment of the corresponding uncertainties is key. As discussed in Section \ref{sec:Temperature Sensor Data Analysis and Modeling} below, we employ considerable effort to accurately estimate those uncertainties in this study.

Using Equation \ref{eqn:vortex_height}, we can probe the central pressure and temperature excursions for a passing vortices using meteorological measurements at the surface, providing an entirely independent assessment of the connection between vortex height and PBL depth. Fortunately, the Mars 2020 Perseverance rover carries high precision, high sampling rate instrumentation ideal for making such measurements. The Mars 2020 Perseverance rover has been operating on the martian surface since landing on 2021 February 18 (at a solar longitude $L_{\rm s} = 5.6^\circ$ -- \citealp{2000P&SS...48..215A}). Although the mission's priorities involve a search for signs of life \citep{2020SSRv..216..142F}, it includes the Mars Environmental Dynamics Analyzer (MEDA) suite. MEDA consists of sensors to measure environmental variables -- air pressure and temperature (the pressure and temperature sensors, PS and ATS respectively), up/downward-welling radiation and dust optical depth, wind speed and direction, relative humidity (via the humidity sensor HS), and ground temperature (via the Thermal Infrared Sensor TIRS). These data enable a wide range of scientific investigation, and periodic public release of the data allows a wide range of participants -- \url{https://pds-geosciences.wustl.edu/missions/mars2020/}.

With a second release, publicly available MEDA data now span nearly 180 mission sols. In a previous study \citep{2022PSJ.....3...20J}, we explored the first 90 sols of data to look for the passage of convective vortices using the PS data, netting 309 encounters. In addition, we sought contemporaneous excursions (positive or negative) in the RDS data to assess whether the vortices were dusty (dust devils) or dustless and found about a quarter of the encounters showed signs of dust lofting. 

For the current study\footnote{As this manuscript was being reviewed, the MEDA team published their own analysis of vortex and dust devil encounters, among other phenomena \citep{Newman2022}. We defer a detailed comparison to a future study.}, we sift the PS data from mission sols 90 through 179 for additional vortex encounters. These data were collected as the local season on Mars shifts from spring to summer. Detailed meteorological modeling of Perseverance's region \citep{2021SSRv..217...20N} predicts a number of seasonal changes, including an increase in vortex activity. Consistent with those predictions, we recovered \numnewvortices\ encounters over 90 sols, almost 70\% more encounters than from the first 90 mission sols. We also use the ATS data to search for temperature excursions that accompany pressure excursions for all encounters from sols 15 through 179 and robustly recover \numvorticeswithDeltaT\ such excursions, allowing us to estimate vortex heights based on Equation \ref{eqn:vortex_height}.

We do not include the MEDA wind data in our analysis here. The first two MEDA data releases did not provide fully processed wind data to correct for, for example, perturbations on the measured winds from the lander body itself \citep{2021SSRv..217...48R}. The third release did include processed winds but only as we were already submitting the present manuscript. Incorporating these data requires significant updates and modifications to our time-series analysis codes, and so we opted to defer that effort to a future study. Many prior studies of martian vortices have used only pressure time series to produce important results, and these studies have employed the assumption that vortex-like pressure dips are vortices. This study follows in that same vein. Moreover, as we show below, our results even without winds still provide significant, novel insights into vortex activity.

In Section \ref{section:Data and Model Analysis}, we provide brief summaries of the MEDA PS and ATS data as relevant to this study. In Section \ref{section:Results}, we discuss our results, including vortex encounter rates and estimates of vortex heights. Finally, in Section \ref{sec:Conclusions}, we summarize conclusions and point out avenues for future study.

\section{Data and Model Analysis} \label{section:Data and Model Analysis}

\begin{figure}
    \centering
    \includegraphics[width=\textwidth]{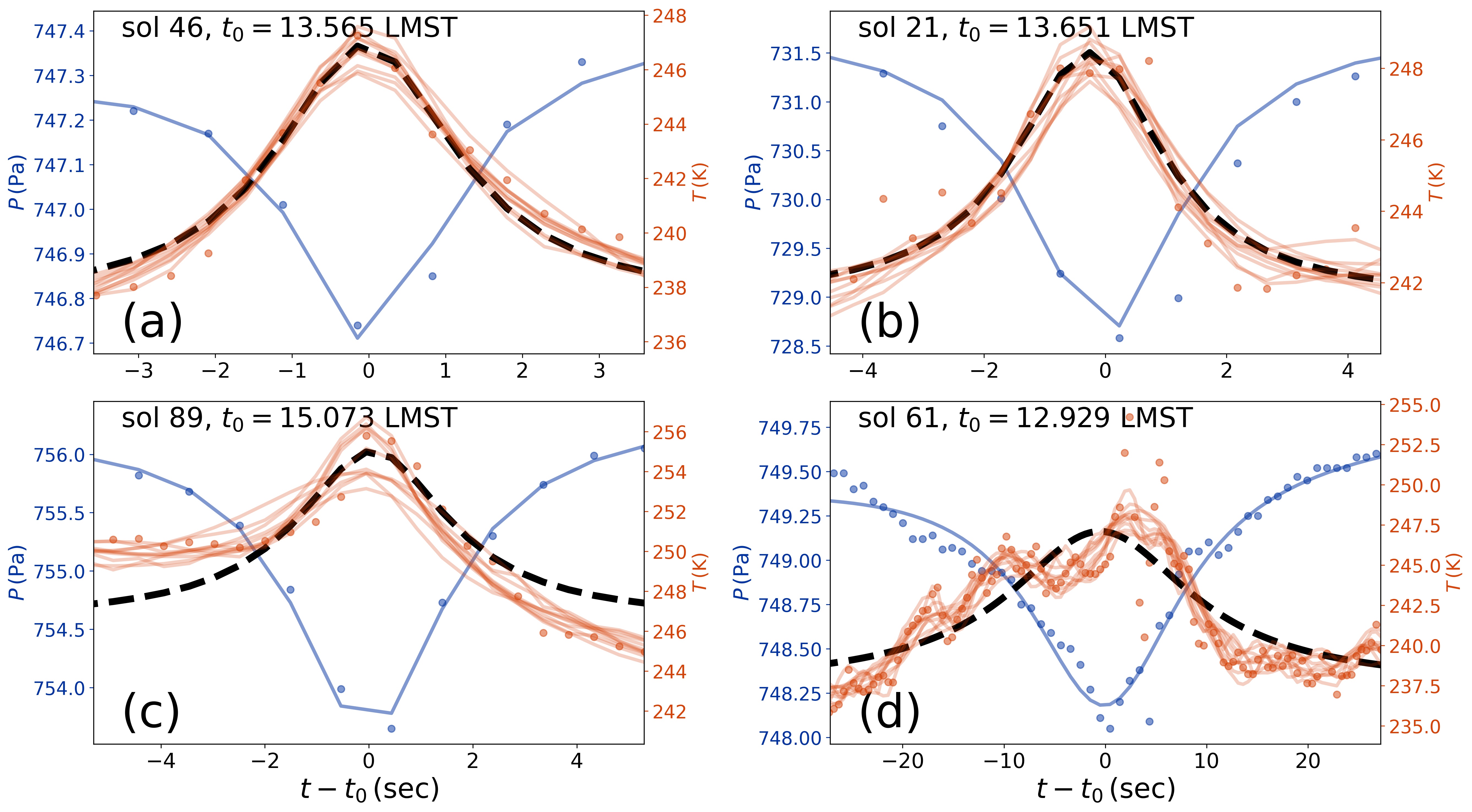}
    \caption{The pressure and temperature signals for several vortices. The blue dots show the pressure signals that were used to find the vortices, while the blue curves show the best-fit pressure models (assuming a white noise model). The orange dots show the contemporaneous temperature data. The dashed black lines show the best-fit temperature models without including any noise, while the orange curves show samples from the modeling that include noise. The sol and central occurrence time $t_0^P$ for each vortex is shown at the top of each panel.}
    \label{fig:temperature_model_figure}
\end{figure}

\subsection{Pressure Sensor Data Analysis and Modeling}

The analysis presented here follows closely the process in \citet{2021PSJ.....2..206J} and \citet{2022PSJ.....3...20J}. We analyzed pressure time series from the PS instrument available from NASA PDS (\url{https://pds-atmospheres.nmsu.edu/PDS/data/PDS4/Mars2020/mars2020\_meda/}), using the data set labeled ``data\_derived\_env''. (The Mars 2020 MEDA PDS Archive Bundle Software Interface Specification provides more details -- \url{https://pds-atmospheres.nmsu.edu/PDS/data/PDS4/Mars2020/mars2020\_meda/document/meda\_bundle\_sis.pdf}.) The data are divided up by mission sol, spanning from midnight one sol to midnight the next, with a sampling rate of $1\, {\rm Hz}$; however, some sols include hour-long (or longer) gaps, and others span only a few hours. Figure \ref{fig:temperature_model_figure} shows example pressure time series for several vortex encounters. 

As in many previous studies \citep[e.g.,][]{2021PSJ.....2..206J}, we use this model for the pressure signal $P(t_i)$:
\begin{equation}
    P(t_i) = B_P + S_P\left( t_i - t_0^P \right) - \frac{\Delta P_0}{1 + \left( \frac{ t_i - t_0^P }{\Gamma_P/2} \right)^2} + \label{eqn:Lorentzian_profile}
\end{equation}
where $B_P$ is the baseline pressure level, $S_P$ is an overall slope, $t_i$ is the time at index $i$, $t_0^P$ is the central time corresponding to the instant of closest approach, $\Delta P_0$ is the pressure excursion amplitude, and $\Gamma_{P}$ is the observed profile full-width/half-max (FWHM) for the pressure signal. (The temperature signals may have different central times and FWHMs, as discussed below.) 

To search for vortex encounters, we first apply a high-pass boxcar filter to the raw pressure data with a window size of \boxcarwindowsize\ to suppress pressure variations that result from meteorological phenomena other than vortices and occur on timescales of hours to days  \citep{2020SSRv..216..148P, 2021SSRv..217...20N}. \citet{2021PSJ.....2..206J} conducted injection-recovery experiments with this scheme to show that this approach sufficiently suppressed the long-term variations without distorting the vortex signals, and the same check on the newly released data analyzed here produced similar results, corroborating this approach.


We then applied a matched filter with a shape given by Equation \ref{eqn:Lorentzian_profile} to the detrended data, subtracting the mean value of the resulting convolution signal $F\ast\Delta P$ and dividing by the standard deviation to scale the spectrum by the intrinsic noise in the dataset. A negative pressure excursion produces a large positive spike in the resulting spectrum. As in \citet{2021PSJ.....2..206J}, we used a threshold value of $F\ast\Delta P \ge$ \detectionthreshold. In setting a detection threshold, other studies \citep[e.g.,][]{Spiga2021} used the pressure excursion itself and required $\Delta P$ exceed some value (usually a few 0.1 Pa). This approach has the advantage of simplicity. However, since the scatter in the pressure time series may vary from sol to sol or hour to hour, this approach may miss excursions or incorrectly identify noise as vortex encounters. In any case, as shown in \citet{2021PSJ.....2..206J}, the threshold $F\ast\Delta P \ge$ \detectionthreshold\ is broadly equivalent to $\Delta P \gtrsim 0.3\,{\rm Pa}$.

Then, for the pressure signals, we fit each putative vortex signal to retrieve best-fit parameters values by using the Levenberg-Marquardt algorithm \citep[cf.][]{Press2007} to maximize the log of the likelihood function $L$: 

\begin{equation}
    \log L = -\frac{1}{2}\sum_{i = 0}^{N-1} \frac{\left( P_i - P(t_i) \right)^2}{\sigma_P^2},\label{eqn:pressure_likelihood}
\end{equation}

where $P_i$ is the pressure value at time $t_i$, $\sigma_P$ is the per-point uncertainty for the pressure time series, and there are $N$ total data points in the vortex signal. (The term on the right-hand is usually called $\chi^2$.) We estimated $\sigma_P$ for each sol by calculating the standard deviation of the boxcar-filtered pressure time series with typical values about $0.07\,{\rm Pa}$. To avoid signal distortion from our detrending process, we fit the original, un-detrended data and estimated uncertainties for the fit parameters ($B_P$, $S_P$, $t_0^P$, $\Delta P_0$, and $\Gamma_P$) as the square root of the corresponding entries in the covariance matrix and scaled by the square root of the reduced $\chi^2$-value for the model fit. This approach is equivalent to imposing a reduced $\chi^2 = 1$ \citep{Press2007}. Example fits for the pressure time series are shown in Figure \ref{fig:temperature_model_figure}.

This process produced a few dozen signals of dubious morphology, so, as in \citet{2022PSJ.....3...20J}, we discarded \discardedvortices\ putative vortex signals with apparent $\Gamma > 250\,{\rm s}$ and $\Delta P_0/\sigma_{\Delta P_0} < 5$ ($\sigma_{\Delta P_0}$ is the uncertainty on $\Delta P_0$). This culling left behind \numnewvortices\ vortices from the newly released MEDA, totaling \totalnumvortices\ with what we call ``credible'' pressure signals when combined with detections from our previous analysis \citep{2022PSJ.....3...20J}.

\subsection{Temperature Sensor Data Analysis and Modeling}
\label{sec:Temperature Sensor Data Analysis and Modeling}
We also analyzed temperature time series from the ATS instrument, also available from NASA PDS as ``data\_calibrated\_env''. These data are also divided up by mission sol and have a sampling rate of $1-2\, {\rm Hz}$. The ATS suite consists of three sensors distributed azimuthally around the rover and two additional sensors on the sides of the rover body near the planet's surface, and each sensor produces a temperature time series labeled by sensor, ATS1 to ATS5. \citet{2021SSRv..217...48R} reported that the temperature sensors have response times shorter than one second, and so we expect that we can recover positive temperature contrasts for the vortices with durations very nearly equal in duration to the negative pressure excursions. 

Some encounters produced temperature excursions on certain sensors and not on others. This variation in the apparent response of the various ATS sensors likely results from a combination of encounter geometry (some vortices likely passed nearer to some sensors than to others) and the complex, local turbulence for each sensor. For simplicity and to avoid the enormous computational demands of applying a Gaussian process analysis (see below) to all five time-series, we selected only a single sensor's time series to analyze for each encounter. We calculated the Pearson correlation coefficient \citep{Press2007} between the pressure and temperature data collected during each encounter and chose the temperature time series with the largest, \emph{negative} correlation on the assumption that the ATS time-series most strongly anti-correlated with pressure was most likely to exhibit a discernible vortex signal. There is no evidence or obvious reason that this approach introduces any biases. 

Having selected the optimal time series, we took temperature data spanning times $t_0^P \pm 3 \Gamma_P$ and fit a temperature excursion model identical in form to the pressure signals: 

\begin{equation}
    T(t_i) = B_T + S_T\left( t_i - t_0 \right) - \frac{\Delta T_0}{1 + \left( \frac{ t_i - t_0^T }{\Gamma_T/2} \right)^2} + \label{eqn:temperature_Lorentzian_profile}
\end{equation}
where, as before, $B_T$ is temperature baseline, $S_T$ is a slope, $\Delta T_0$ is the temperature excursion amplitude, $t_0^T$ is the central encounter time, and $\Gamma_T$ is the FWHM for the temperature signal. 

Given the unique properties of the temperature data, we used a different approach to analyzing them than we used for the pressure data. We did not use the temperature data to detect vortices. Instead, we used the times for encounters derived from the pressure data and then analyzed the corresponding times in the optimal ATS time series to search for thermal excursions that coincided closely in time. The ATS data showed considerably more turbulent fluctuations than the PS data, likely as the result of near-surface turbulence (see panel (d) of Figure \ref{fig:temperature_model_figure}). Though the vortex-associated temperature signals can often be prised from the grip of this non-white noise, robust assessment of the signal requires a more complex noise model -- a Gaussian process model \citep{books/lib/RasmussenW06, 2018RemS...10...65J}.

To fit Equation \ref{eqn:temperature_Lorentzian_profile} to each encounter, we employed the following likelihood function for the temperature data:
\begin{equation}
    \log L = -\frac{1}{2} \vec{r}^{T}\ \mathbf{K}^{-1}\ \vec{r} - \frac{1}{2} \ln \mathrm{det}\ \mathbf{K} - \frac{N}{2} \ln\left(2\pi\right),\label{eqn:temperature_likelihood}
\end{equation}
where $\vec{r}$ is the vector of residuals between the model and data points and $\mathbf{K}$ is the covariance matrix. In traditional least-squares minimization schemes (as for the Levenberg-Marquardt scheme used for the pressure data), the covariance matrix is assumed to be diagonal, i.e. $\mathbf{K_{ij}} = \sigma_i^2 \delta_{ij}$, where $\sigma_i$ is the per-point uncertainty and $\delta_{ij}$ is the Kronecker $\delta$ ($\delta_{ii} = 1$ and $\delta_{i \ne j} = 0$). This approach is equivalent to assuming the scatter for each data point is independent of all the other points. By contrast, for a Gaussian process approach, the covariance matrix is allowed to have off-diagonal elements whose function form is prescribed, i.e. $\mathbf{K_{ij}} = \sigma_i^2 \delta_{ij} + k(t_i, t_j)$ where $k(t_i, t_j)$ is the kernel function relating data points at times $t_i$ and $t_j$. A variety of choices for $k$ is possible, with some implications and limitations arising from the different choices \citep{books/lib/RasmussenW06}. For our analysis, we chose a form representing a stochastically-driven, damped harmonic oscillator, which opens the door to some optimization schemes to reduce compute times \citep{celerite1, celerite2}. 

We used a Markov chain-Monte Carlo (MCMC) sampler (emcee -- \citealp{2013PASP..125..306F}) to map the posterior distributions for the best-fit parameters $B_T$, $S_T$, $\Delta T_0$, $t_0^T$, and $\Gamma_T$ (plus four parameters for the kernel function). We used 32 walkers, a burn-in chain of 5,000 links, and then production chains of 10,000 links and checked convergence of the chains by requiring autocorrelation times at least 20 times shorter than the total chain lengths (usually, the chains were more than 50 times longer than the autocorrelation times) for all the fit parameters -- see \citet{2013PASP..125..306F} for details about this approach.

\begin{figure}
    \centering
    \includegraphics[width=\textwidth]{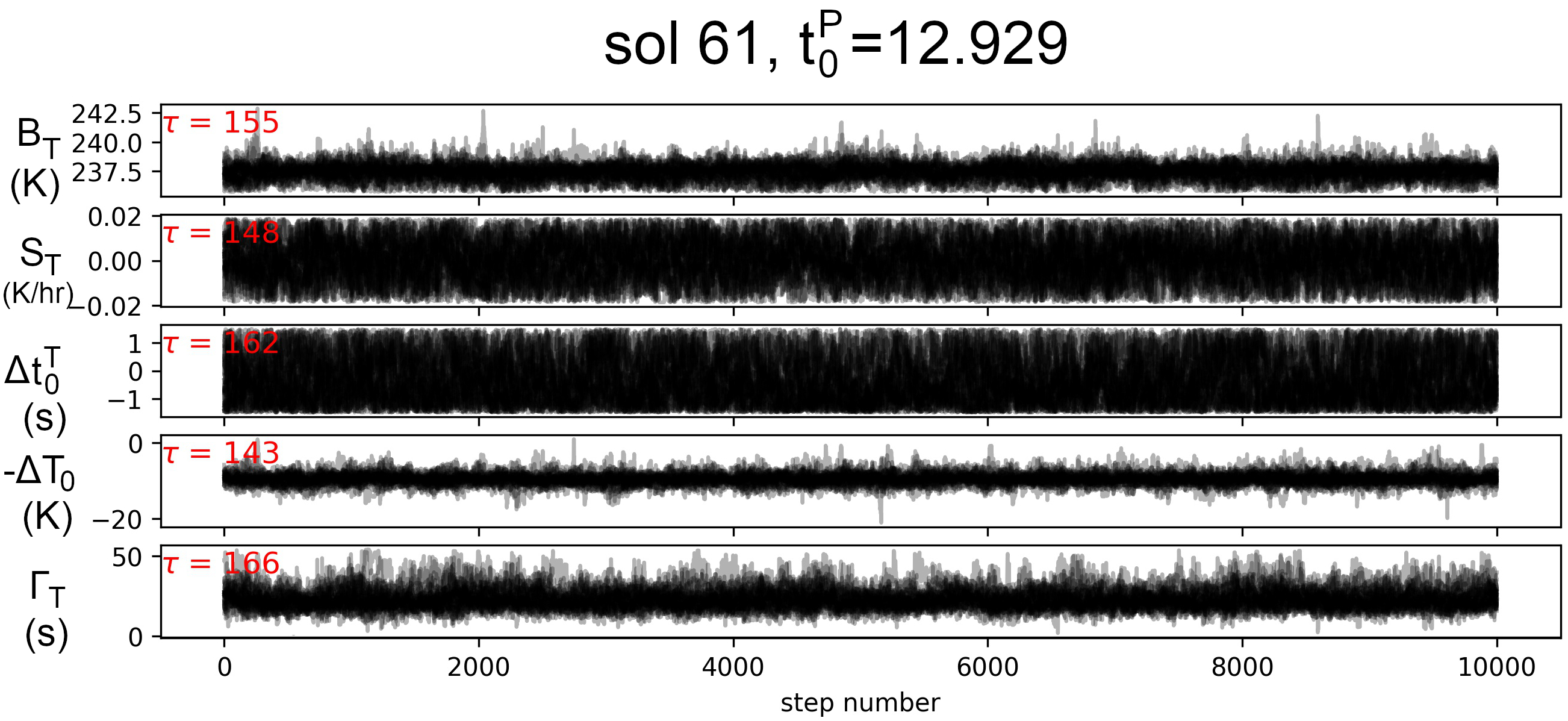}
    \caption{MCMC chains for the fit parameters for the temperature signal in Figure \ref{fig:temperature_model_figure} (d). The autocorrelation times $\tau$ are shown in red. (We fit $-\Delta T_0$ and the difference between the central encounter times for the pressure and temperature signals $\Delta t_0^T$.)}
    \label{fig:MCMC-example_sol61-vortex1}
\end{figure}

\begin{figure}
    \centering
    \includegraphics[width=\textwidth]{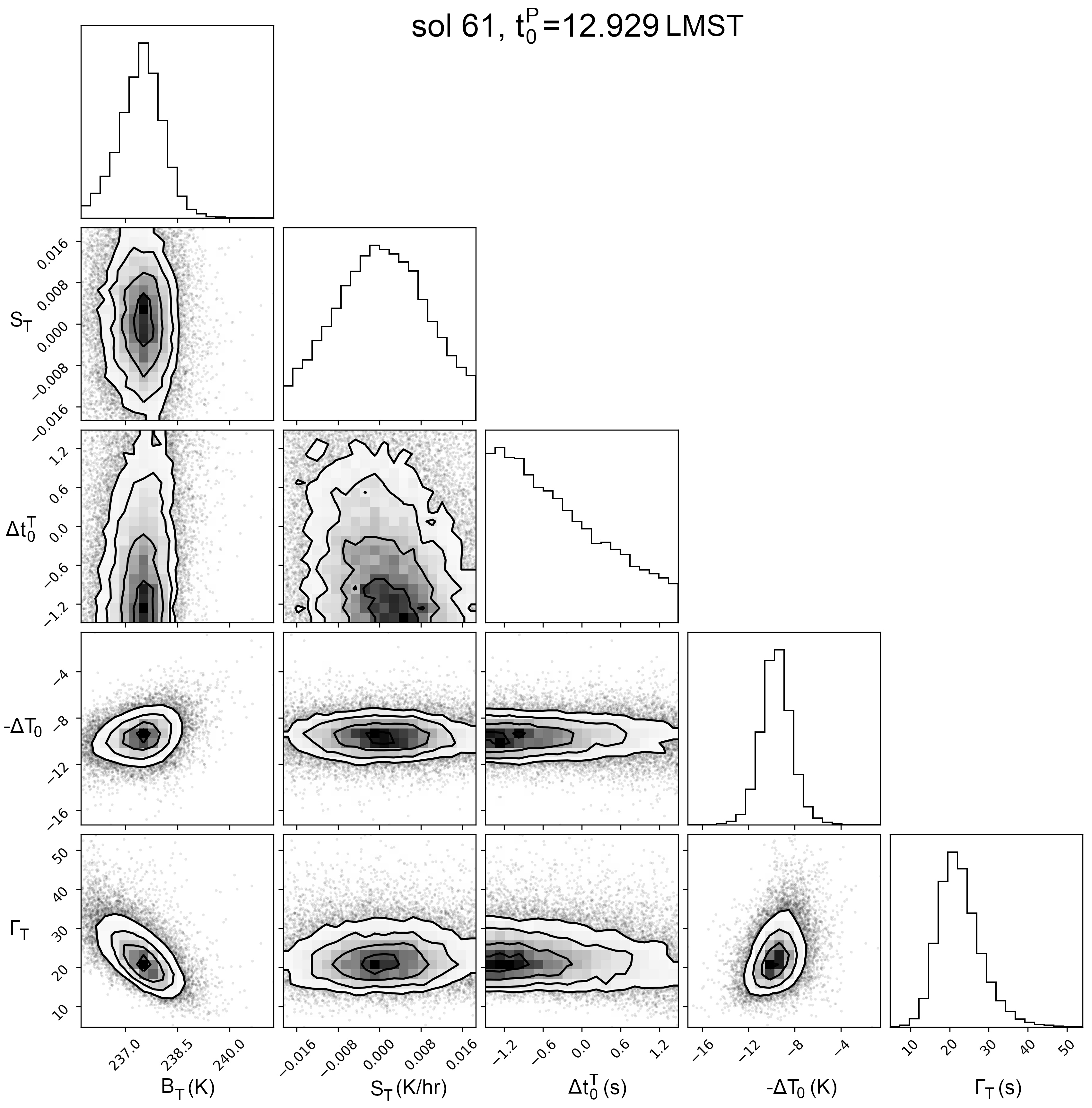}
    \caption{The marginalized posterior distributions for the fit parameters from Figure \ref{fig:MCMC-example_sol61-vortex1} and for the temperature model in Figure \ref{fig:temperature_model_figure}(d).}
    \label{fig:posteriors-example_sol61-vortex1}
\end{figure}

Figure \ref{fig:MCMC-example_sol61-vortex1} shows an example of the production chains for fitting the temperature signal shown in Figure \ref{fig:temperature_model_figure}(d). By-eye inspection shows the chains effectively exploring the solution space, and convergence of the chains is suggested by the autocorrelation time for each chain. Figure \ref{fig:posteriors-example_sol61-vortex1} shows the marginalized posterior distributions corresponding to the MCMC chains from Figure \ref{fig:MCMC-example_sol61-vortex1}. 

As is typical for credible detections, the posterior for $\Delta T_0$ shows the signal is inconsistent with $\Delta T_0 = 0$, meaning the excursion is robustly detected. Other parameters have less robust constraints. For example, the posterior for the central encounter time (shown as $\Delta t_0^T$ in Figures \ref{fig:MCMC-example_sol61-vortex1} and \ref{fig:posteriors-example_sol61-vortex1}) spans the full allowed solution space. Inspection of Figure \ref{fig:temperature_model_figure}(d) illustrates why: the non-white noise masks the central peak. Anticipating this difficulty, we constrained the solution space for $\Delta t_0^T$ to lie within $\pm$ five times the error bars for $t_0^P$. For other fit parameters, we allowed much broader solution spaces. The $\Gamma_T$ values, for instance, were allowed to span the entire time span for each vortex encounter. For uncertainties on each of the temperature signal parameters, we took the 16\% and 84\% quartiles for the posteriors, the equivalent of $\pm1$-standard deviation for a Gaussian distribution.

As with our pressure signals, we culled the list of temperature signals using the following criteria:
\begin{enumerate}
    \item MCMC chains must have converged -- As described above, this criterion is met by requiring an autocorrelation time at least 20 times shorter than the chain length. Most converged chains have even shorter autocorrelation times.
    \item Vortices must have $\Delta T_0 / \sigma_{\Delta T_0} >$ \DeltaTSNR\ -- This criterion ensures a non-zero, positive amplitude.
    \item The best-fit $t_0^T$ must agree to within uncertainties with the best-fit $t_0^P$ value -- This criterion means a temperature excursion must coincide in time with the pressure excursion. Otherwise, the temperature signal may simply arise from turbulent noise. (Uncertainties for $t_0^P$ and $t_0^T$ were added in quadrature.)
    \item The $\Gamma_T$ posterior must not span the whole allowed range for the solution space -- Posteriors that are too wide provide no meaningful constraint on the signal's duration.
    \item The best-fit $\Gamma_T$ must agree to within uncertainties with the best-fit $\Gamma_P$ value -- This criterion makes it more likely that the putative temperature signal corresponds to a true vortex encounter. (Again, uncertainties for $\Gamma_P$ and $\Gamma_T$ were added in quadrature.)
\end{enumerate}

From among the \totalnumvortices\ vortices with credible pressure signals from sols 15 through 179, these criteria leave \numvorticeswithDeltaT\ vortices with what we call ``credible'' temperature signals. Vortex encounters that failed to produce a credible temperature signal often fail to satisfy multiple criteria at once (e.g., the best-fit excursion has a $t_0^T$ value inconsistent with the corresponding $t_0^P$ and the $\Gamma_T$ posterior spans the whole allowed range), and there does not seem to be any one overwhelmingly common failure mode. There also does not seem to be a pattern for which encounters are most likely to fail \emph{except} for the encounters with the smallest $\Delta P_0$ values. However, this latter trend is entirely consistent with expectations: small $\Delta P_0$ corresponds to small $\Delta T_0$, the signals for which are more likely to be swallowed by turbulent noise. Exploring how this bias might skew the inferred vortex heights discussed in Section \ref{section:Results}, we leave for future work.

%
%

\section{Results} \label{section:Results}

\begin{figure}
    \centering
    \includegraphics[width=\textwidth]{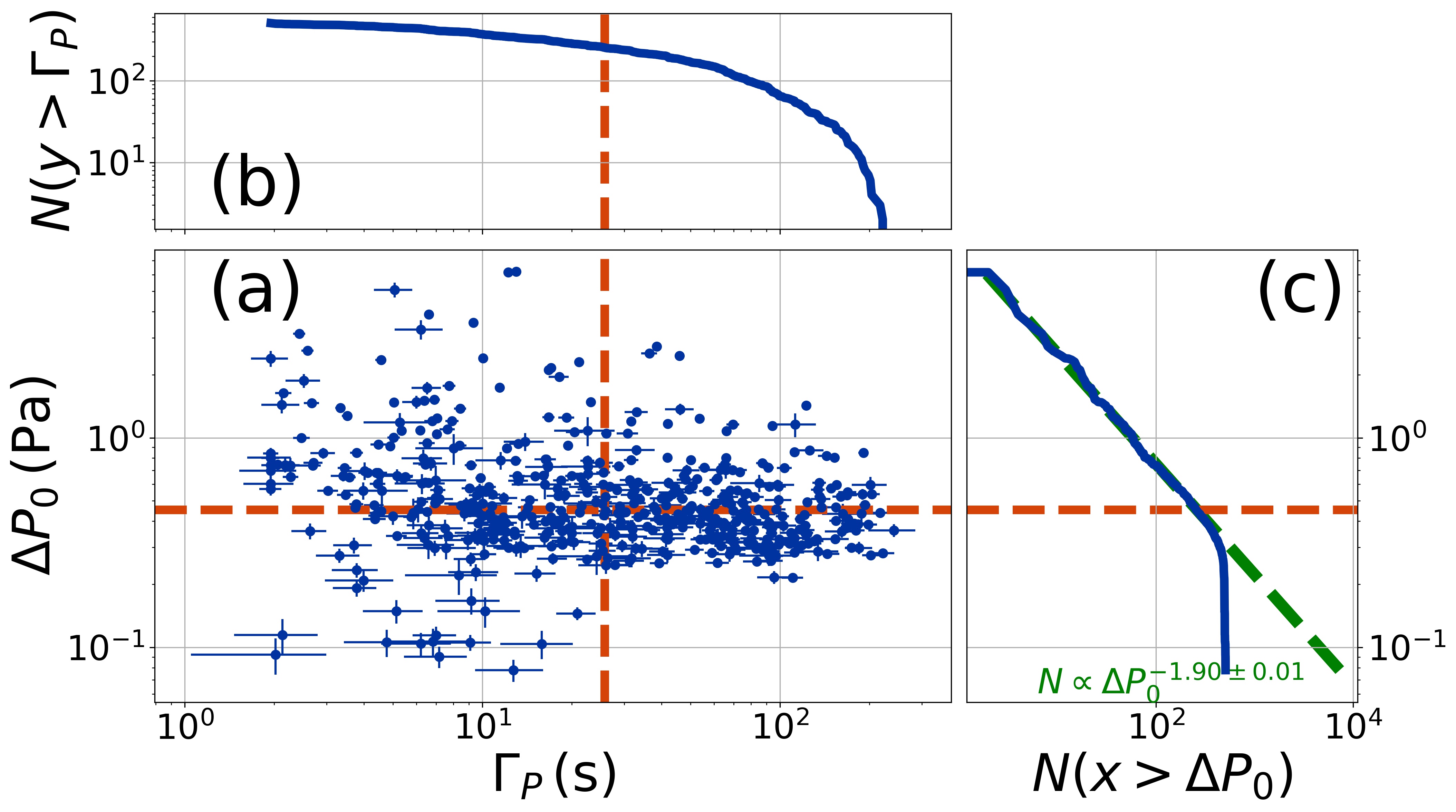}
    \caption{(a) The best-fit $\Delta P_0$- and $\Gamma_P$-values (blue dots) with error bars for sols 90 through 179. (b) Cumulative histogram of $\Gamma$-values, along with the median value ($\Gamma = \left( 19.7 \pm 1.6 \right)\,{\rm s}$) shown by the dashed, orange line. (c) Cumulative histogram of $\Delta P_0$-values, along with the median value ($\Delta P_0 = \left( 0.45 \pm 0.01 \right)\,{\rm Pa}$) shown by the dashed, orange line. The dashed green line shows a power-law fit to the histogram for $\Delta P_0 > 0.45\, {\rm Pa}$ with $N \propto \Delta P_0^{-1.90\pm0.01}$.}
    \label{fig:DeltaP-Gamma_scatterplot-histogram}
\end{figure}

Figure \ref{fig:DeltaP-Gamma_scatterplot-histogram} illustrates the resulting best-fit $\Delta P_0$- and $\Gamma_P$-values for the \numnewvortices\ vortices from sols 90 through 179. Because of the difficulty to detect vortices with small $\Delta P_0$, we fit a power-law to the cumulative histogram of $\Delta P_0$-values only for values above the median, $\Delta P_0 = 0.45\, {\rm Pa}$, using the Levenberg-Marquardt algorithm. Poisson sampling was assumed to estimate bin uncertainties. The histogram knee reported in previous studies \citep{Spiga2021} is not apparent in these data. The best-fit power-law index here is roughly consistent with previous work. We found an index of $-1.99\pm0.02$ in our previous study of PS data from sols 15 through 89 \citep{2022PSJ.....3...20J}. The results also comport with studies of datasets from other missions. \citet{2021Icar..35514119L} considered vortex pressure signals from the InSight Mission and, considering vortices with $0.8\,{\rm Pa} < \Delta P_0 < 3\,{\rm Pa}$, found a power-law index of $-2$. 

\begin{figure}
    \centering
    \includegraphics[width=\textwidth]{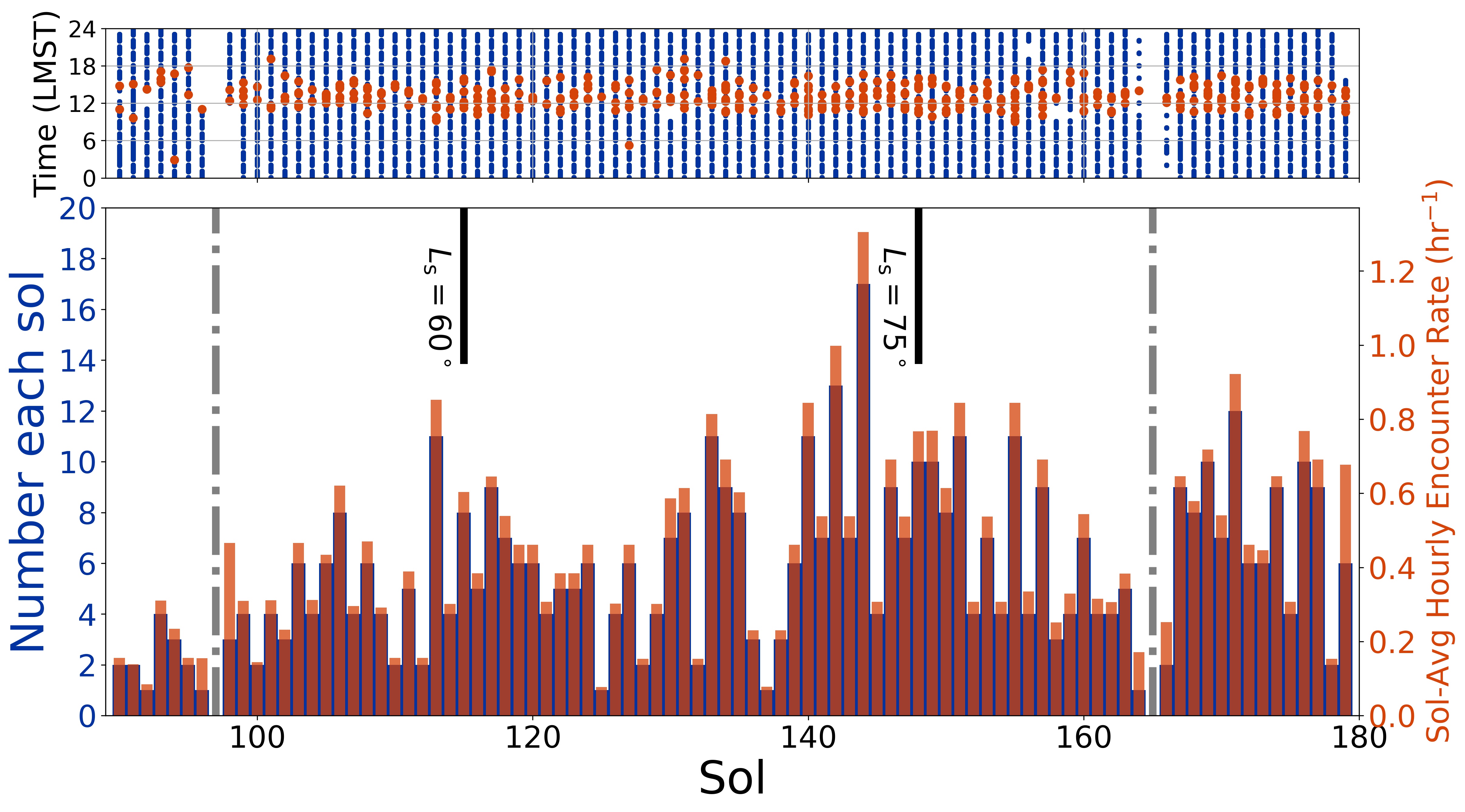}
    \caption{(Top) When pressure data were collected during each sol. The orange dots also show when vortex signals were detected. (Bottom) The blue bars show the number of vortex encounters during each sol, while the orange bars show the number of encounters each sol divided by the total number of hours during which the pressure sensor collected data. The grey dash-dotted lines show sols when data were unavailable. Solid black lines indicate $L_{\rm s}$-values. }
    \label{fig:vortex_occurrence_each_sol}
\end{figure}

\begin{figure}
    \centering
    \includegraphics[width=\textwidth]{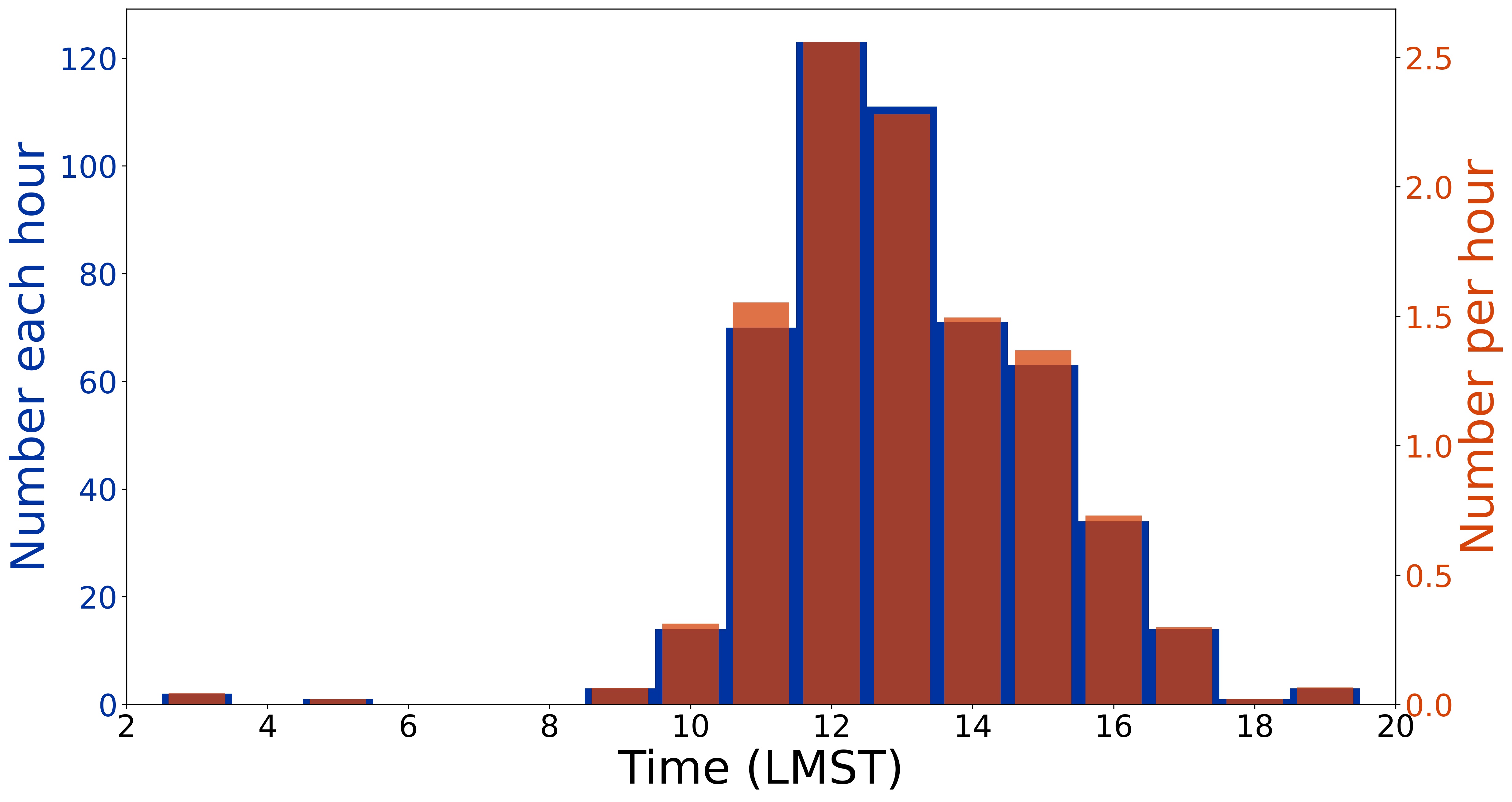}
    \caption{The blue bars show the total number of vortex encounters that took place during that hour over the whole 90 sol period, while the orange bars show that number divided by the total number of hours that hour of the day was observed.}
    \label{fig:vortex_occurrence_each_hour}
\end{figure}

Based on these detections, we can also assess the vortex encounter rates. Figure \ref{fig:vortex_occurrence_each_sol} bins the encounters by sol. (Sols 97 and 165 have no data available.) These results show that there was an average rate of $5.8\pm2.4$ encounters per sol (where error bars come from assuming Poisson statistics). Some sols had more data available than others, however, and so we divided the number of encounters on a given sol by the total number of hours of observational data, giving the orange bars in Figure \ref{fig:vortex_occurrence_each_sol}. The mean for all the sol-averaged hourly encounter rates is $0.5\pm0.2\,{\rm hr^{-1}}$, with variations between 0.1 and $1.3\,{\rm hr^{-1}}$. Some of that variability is likely meteorological, and some is due to variations in when, during each sol, data were collected (see the top panel of Figure \ref{fig:vortex_occurrence_each_sol}).

Encounter rates also show hour-to-hour variation, as illustrated in Figure \ref{fig:vortex_occurrence_each_hour}. We have normalized the number of encounters during each hour by the total number of hours (over all available/usable sols) to estimate the hourly encounter rate. The encounter rate peaks about mid-day, in this case at 2.6$\pm 0.2$ encounters per hour (once every 20 minutes), dropping below detectable levels early in the morning and late in the afternoon (although there were a handful of very early morning encounters).

The trends in encounter rates are generally consistent with results from our previous analysis of sols 15 through 89 vortex encounters -- a peak in the hourly rate near mid-day with a quicker ramp up in the morning than the fall-off in the afternoon. One big difference, though: the maximum encounter rates reported here exceed the maximum from our previous study by more than 70\%. This result corroborates predictions from \citet{2021SSRv..217...20N} that vortex activity should increase going into summer solstice. A complete comparison with results of \citet{2021SSRv..217...20N} requires converting the encounter rates into occurrence rates \citep[cf.][]{2021Icar..35814200K}, but these results are at least consistent with an increased occurrence rate.

We can assess the encounter rates by estimating the fractional area covered by dust devils by comparing the total durations of all vortex encounters to the total observational time \citep{2021Icar..35514119L}. The total duration of encounters reported here is 6.4 hours, while the total observational time is about 1118.7 hours, giving a fractional area of 0.57\% compared to 0.37\% for the previous sol 15 to 89 dataset. This result suggests again that the recent vortex occurrence rate is significantly larger.

\begin{figure}
    \centering
    \includegraphics[width=\textwidth]{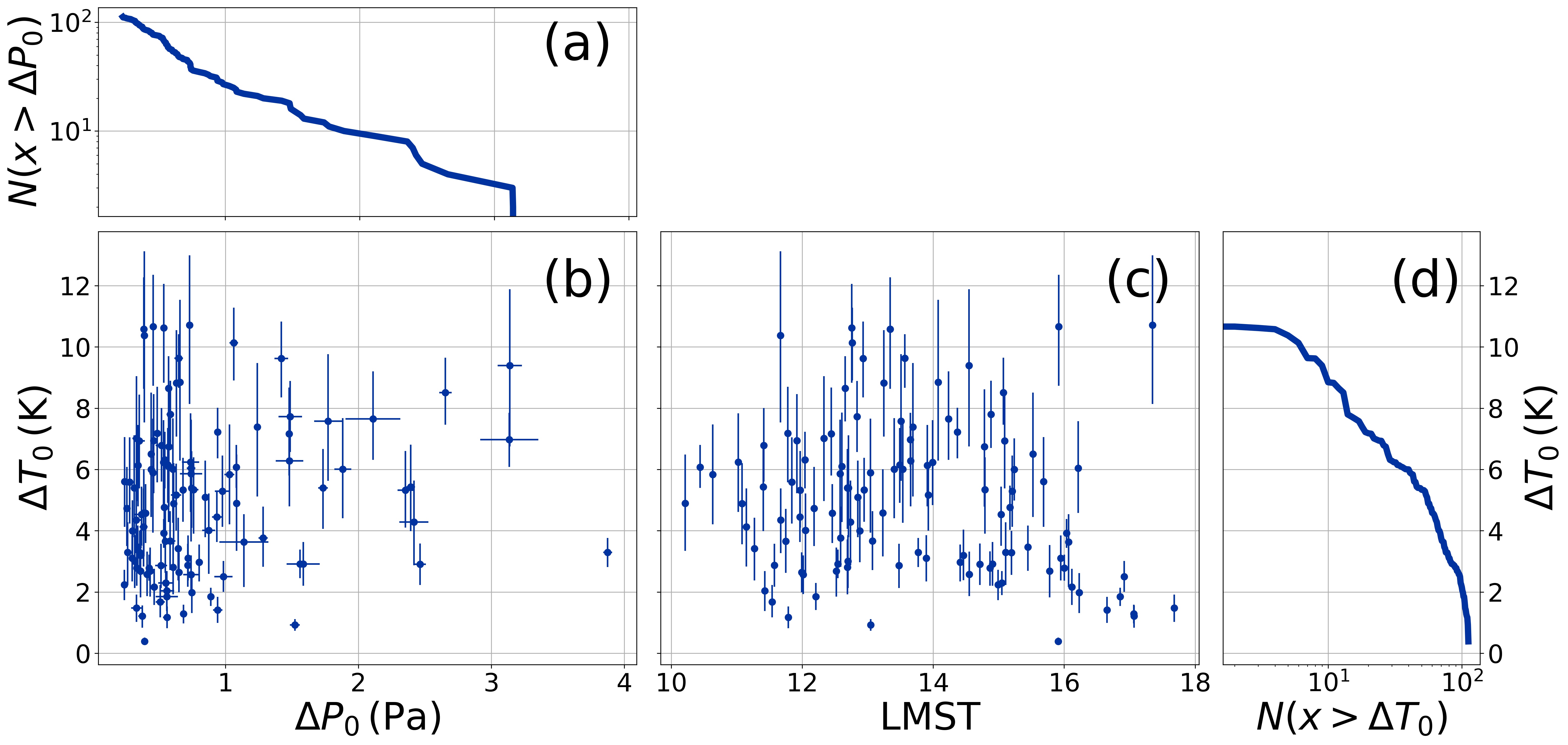}
    \caption{Estimated temperature signal amplitudes $\Delta T_0$ vs.~the pressure signal amplitudes $\Delta P_0$ (b) and vs.~time of day LMST (c) for the \numvorticeswithDeltaT\ vortices with credible temperature signals. Histograms of $\Delta P_0$ and $\Delta T_0$ for the vortex encounters with credible temperature excursions are shown in (a) and (d), respectively.}
    \label{fig:DeltaT0_vs_DeltaP0_and_vs_LMST}
\end{figure}

Turning to the temperature signals, Figure \ref{fig:DeltaT0_vs_DeltaP0_and_vs_LMST} shows the distribution of $\Delta T_0$ values for the \numvorticeswithDeltaT\ vortices with credible detections (along with associated error bars). The largest $\Delta T_0$ value is $10.7^{+2.6}_{-2.3}\,{\rm K}$, the median value $4.9\pm1.6\,{\rm K}$, and the minimum value $0.4\pm0.1\,{\rm K}$. These values compare favorably with the small number of detections of vortex thermal signals from other surveys, martian and terrestrial. As part of a re-analysis of Viking Lander 2 meteorological data, \citet{2003Icar..163...78R} reported detections of 38 vortex encounters, several with accompanying temperature excursions with amplitudes between 0 (i.e., no clear detection) and $6\pm2{\rm K}$. Resolution of those signals was particularly challenging with that dataset since the highest sample rate was once every $8\,{\rm s}$ ($0.125\,{\rm Hz}$). \citet{2000JGR...105.1859R} surveyed data from the Mars Pathfinder ASI/MET instrument and reported 19 vortex encounters, estimating temperature signal amplitudes via a scaling theory \citep{1998JAtS...55.3244R} from 0.6 to $2.1\,{\rm K}$.

Figure \ref{fig:DeltaT0_vs_DeltaP0_and_vs_LMST}(a) plots the temperature excursions against the corresponding $\Delta P_0$ values, and a subdued positive correlation appears. Pearson's r correlation coefficient for the two distributions is $0.15$ but with a very marginal $p$ value of about 10\% (meaning there is a 10\% chance for such a correlation coefficient even if the two distributions were drawn at random). Panel (b) shows $\Delta T_0$ vs.~time of day, where any correlation appears even less pronounced, though still potentially present. Morning encounters seem to skew toward smaller $\Delta T_0$, perhaps consistent with the expectation that vortex activity is muted by a less vigorous boundary layer \citep{2011RvGeo..49.3005P}. Encounters near mid-day, when vortex activity is at its height, span a wide range of values, though perhaps with smaller values less represented. Finally, later in the day, activity peters out, and the bulk (such as it is) of encounters have smaller $\Delta T_0$. Even so, the late afternoon/early evening still sees two of the largest $\Delta T_0$ encounters. The limits of small number statistics very likely bear on these patterns, though, and additional detections as the Mars 2020 mission continues may clarify correlations. 

\begin{figure}
    \centering
    \includegraphics[width=\textwidth]{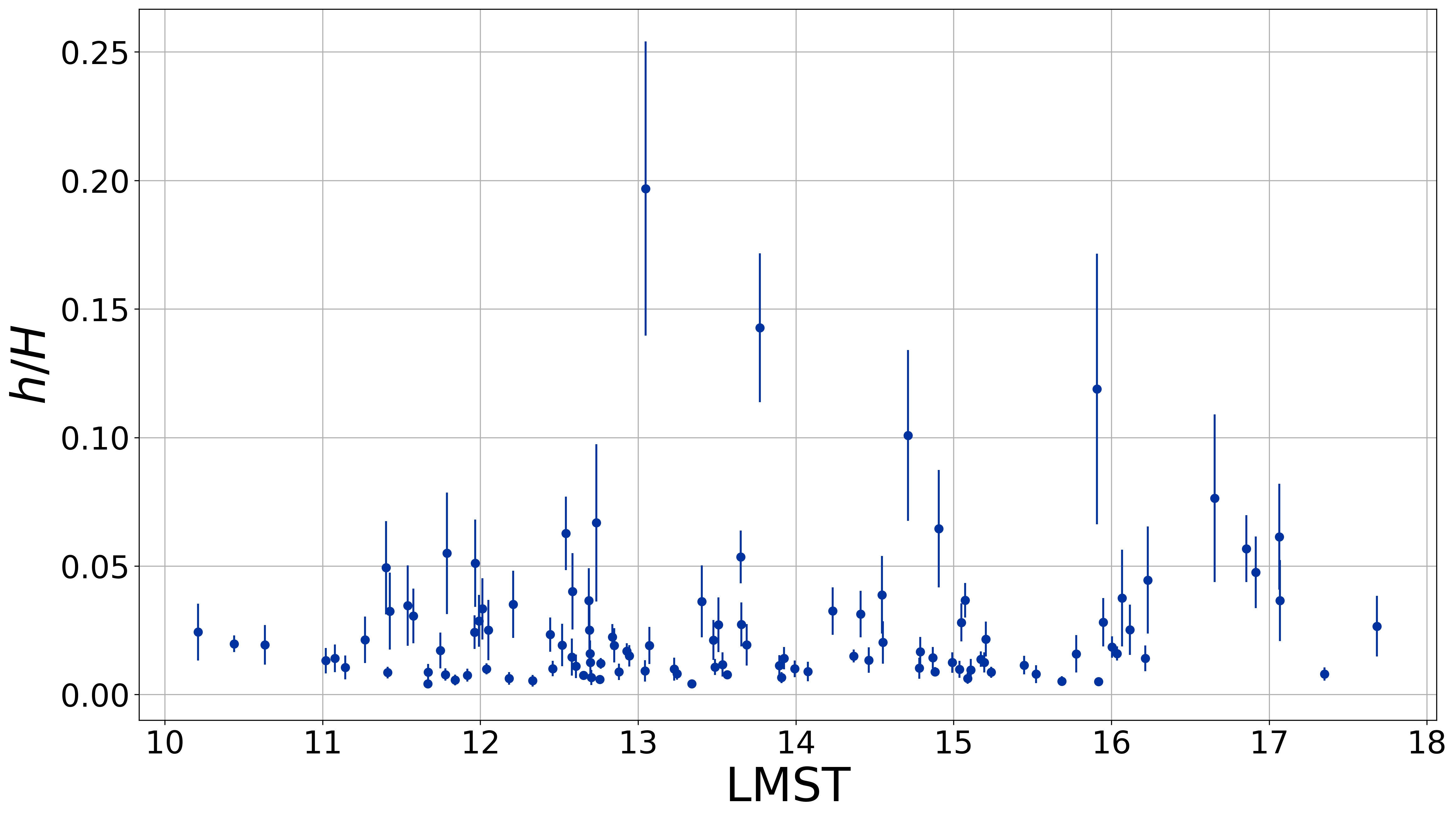}
    \caption{Estimates for the heights $h$ of the encountered convective vortices, scaled by the atmospheric scale height $H$ as a function of time of day LMST. Error bars derive from the uncertainties on the vortex $\Delta P_0$ and $\Delta T_0$ values.}
    \label{fig:h_over_H}
\end{figure}

Finally, using Equation \ref{eqn:vortex_height}, we can convert our best-fit $\Delta P_0$ and $\Delta T_0$ values into estimates for the height of the convective vortices, shown in Figure \ref{fig:h_over_H}. Uncertainties for $h/H$ incorporate all uncertainties on the relevant fit parameters as depicted in Equation \ref{eqn:vortex_height}. The largest value for $h/H$ is $0.20\pm0.06$, the median value $0.027\pm0.006$, and the smallest value $0.004\pm0.001$. Taking the martian atmospheric scale height as $11\,{\rm km}$ \citep{2011RvGeo..49.3005P}, these translate to $2.2\,{\rm km}$, $300\,{\rm m}$, and $40\,{\rm m}$ respectively. 

As for other results here, small number statistics also likely mask trends in Figure \ref{fig:h_over_H} as well, but the trends that seem apparent comport with expectations. Dust devils occurring during any given period of the day span a range of heights and diameters, and evidence suggests that diameters follow a power-law with an index of -2 \citep{2011Icar..215..381L} (and that a dust devil's height scales with diameter squared -- \citealp{2020Icar..33813523J}). Since the PBL height likely sets an upper limit to vortex height \citep{2015Icar..260..246F}, we might expect to see a range of vortex heights during any period of the day on Mars but with the tallest vortices occurring in the middle of the day when the PBL is deepest \citep{2001imm..book.....A, 2013JGRE..118.1468C}. Indeed, as shown in Figure \ref{fig:h_over_H}, $h/H$ is confined to relatively small values early in the day, after which the largest value in any time span grows toward mid-day before declining into late afternoon and evening. However, PBL depth is thought to peak around 15:00 or 16:00 local time, depending on the exact model parameters and observational data considered, later than the time of occurrence of our $h/H$ maximum. Whether that disagreement arises from the small number statistics for our results is unclear, though. In any case, our vortices do not seem to ascend to the top of PBL. During the $L_{\rm s}$ values when our observations were collected, estimates suggest the PBL reaches between 5 and $6\,{\rm km}$, while our vortices tap out at about $2.2\,{\rm km}$. Auspiciously, \citet{2015Icar..260..246F} found that their heights typically reached about 1/5 of the PBL height (1 to $1.2\,{\rm km}$), consistent with the largest $h/H$ values we report.

\begin{figure}
    \centering
    \includegraphics[width=\textwidth]{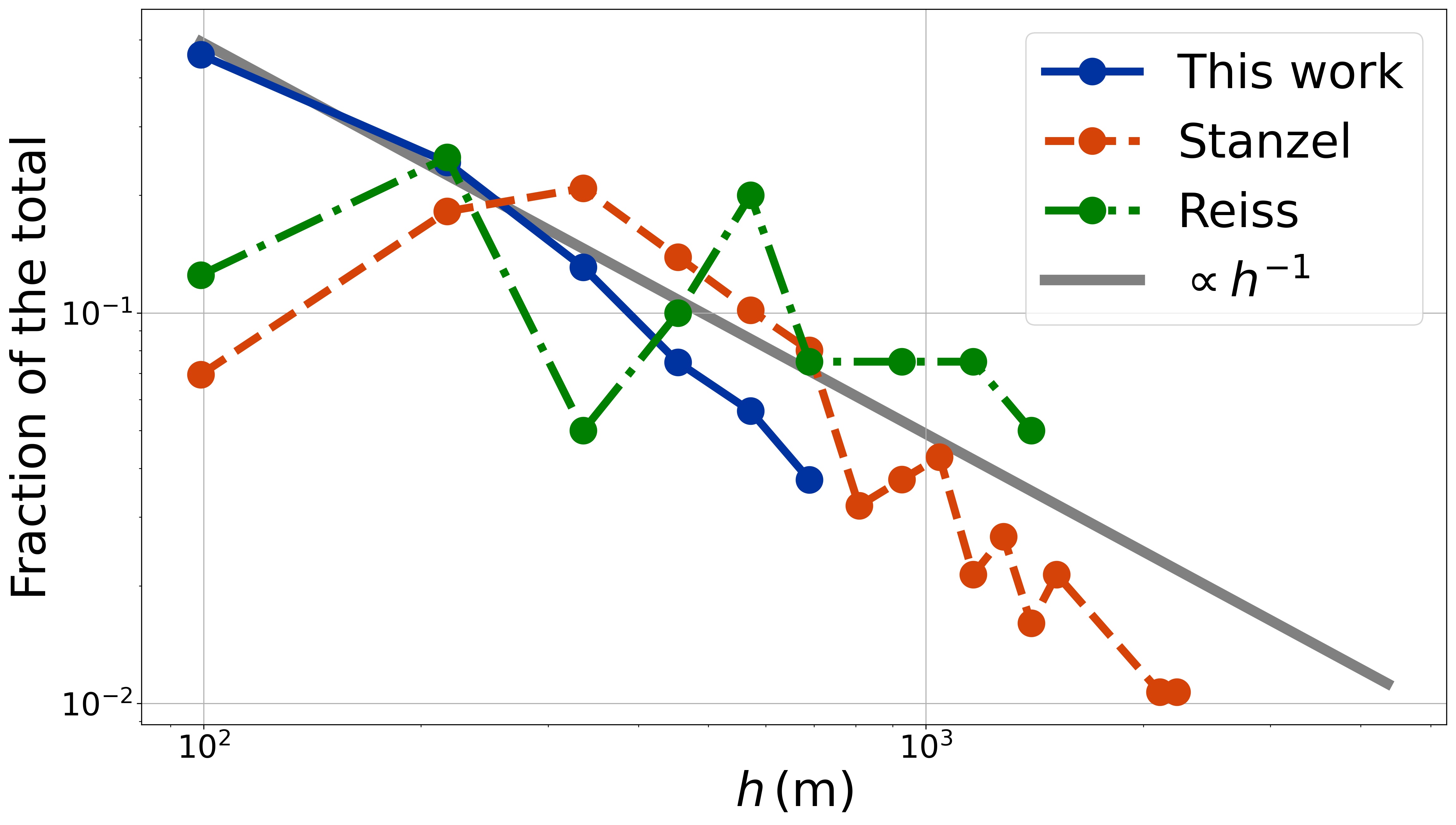}
    \caption{Comparison of the vortex heights estimated from this work (blue, solid curve) to those from \citet{2008Icar..197...39S} (orange, dashed curve) and \citet{2014Icar..227....8R} (green, dash-dotted curve). To guide the eye, a curve showing a $\propto h^{-1}$ dependence (grey, solid) is included. The y-axis shows the fraction of that dataset that falls within that bin. For instance, about 46\% of the vortices reported in ``This work'' fall into the smallest bin, centered on $100\,{\rm m}$.}
    \label{fig:comparing_to_Stanzel-and-Reiss}
\end{figure}

We can also compare our heights to those estimated from image surveys of dust devils. \citet{2008Icar..197...39S} measured heights and diameters of 205 dust devils using High Resolution Stereo Camera (HRSC) images from the ESA Mars Express orbiter, images with resolutions of $12.5\, {\rm m\ pixel^{-1}}$. Devils had diameters from $45\,{\rm m}$ to $1\,{\rm km}$ with a median of $63\,{\rm m}$ and heights from $75\,{\rm m}$	to $3.4\,{\rm km}$ and a median of $450\,{\rm m}$. \citet{2014Icar..227....8R} measured heights and diameters of 47 dust devils using imagery from the Mars Reconnaissance Orbiter's (MRO) Compact Reconnaissance Imaging Spectrometer for Mars (CRISM), the Context Camera (CTX), and High Resolution Imaging Science Experiment (HiRISE), resolutions between 0.25 and $18\,{\rm m\ pixel^{-1}}$. For that study, diameters spanned 15 to $280\,{\rm m}$ with a median of $75\,{\rm m}$, while heights spanned $40\,{\rm m}$ to $4.4\,{\rm km}$ with a median of $530\,{\rm m}$. Figure \ref{fig:comparing_to_Stanzel-and-Reiss} compares the distributions of heights from these studies to those for our detected vortices. (Unfortunately, \citet{2015Icar..260..246F} did not report their results in table form, though the histogram of heights therein peaked at about $500\,{\rm m}$.) The dust devils imaged were typically much taller than our vortices, but the limited pixel resolution very likely contributes to a dearth of detections in the smallest bins. Disagreement between the histogram peaks aside, reassuringly, they all fall off with a power-law index statistically consistent with -1, suggesting they capture similar behavior, at least the population-level, though small number statistics certainly play a role. 

A power-law index of -1 is not entirely consistent with previous work that suggests that (1) the distribution $p(D)$ of dust devil diameters $D$ falls off roughly as $D^{-2}$ \citep{2011Icar..215..381L} and (2) $D \propto h^{1/2}$ \citep{2020Icar..33813523J}. Assuming $D \propto h^{1/2}$, we can convert $p(D) \propto D^{-2}$ to a distribution in $h$ as $p(h) = p(D) \times \partial D/\partial h \propto h^{-3/2}$. However, the power-law indices for the Stanzel and ``This work'' datasets in Figure \ref{fig:comparing_to_Stanzel-and-Reiss} \emph{are} consistent (at $3 \sigma$) with $-3/2$ as well (the index for the Reiss dataset is not consistent with -3/2). An enlarged sample of vortex heights may help clarify this relationship.

\section{Discussion and Conclusions}
\label{sec:Conclusions}

So far as we are aware, this study is the first to estimate vortex height using measurements of meteorological time series, and our approach opens the door to several other important investigations of boundary layer models and processes. Our application of Gaussian process analysis enables a robust estimate of temperature excursions associated with vortices, and comparing these with the wind field associated with a vortex enables a potentially transformative test of the bedrock model of vortex thermodynamics \citep{1998JAtS...55.3244R}. As popular as this model has been for decades, it remains essentially untested. A definitive test would substantially bolster the usefulness of dust devils and vortices as probes of PBL processes. 

A comprehensive assessment of all meteorological variables associated with dust devils could also be transformative for understanding their role in erosion and atmospheric dust. The exact contribution of dust devils to the martian dust budget, for instance, remains highly uncertain, with estimates spanning from a negligible to a dominant source \citep{2016SSRv..203...89F}. MEDA's data promise to be especially enlightening as they provide constraints on vortex pressure and wind fields, perturbations to insolation associated with light scattering by dust (reflective of the dustiness of a vortex), and, as shown here, temperature excursions. Potentially, combining winds and dust content for vortex encounters would directly establish the relationship between winds and dust lifting capacities that has remained elusive \citep{2006RvGeo..44.3003B}. MEDA wind measurements, combined with temperature fluctuations, allow estimates of sensible heating rates \citep{Spiga2021}. The model of \citet{1998JAtS...55.3244R} makes specific predictions relating vertical wind speeds within vortices to heating rates, and MEDA's wind instrument measures vertical as well as horizontal winds \citep{2021SSRv..217...48R}, providing a direct test of the model. Combining vortex dust content with vertical winds would also give direct estimates of dust fluxes, and adding in estimated vortex heights would give a sense for the altitudes to which dust is delivered. In addition, comparing wind speeds and temperature fluctuations within vortices would also help constrain engineering models for maximum expected wind speeds \citep{2021Icar..35414062L}. 

The results reported here come with some important caveats. Stationary meteorological sensors have a very small chance to pass directly through the center of a vortex and therefore likely probe conditions some distance from the center -- the so-called ``miss distance effect'' \citep{2014JAtS...71.4461L, 2018Icar..299..166J, 2020Icar..33513389K}. Consequently, the observed pressure and temperature excursions, $\Delta P_0$ and $\Delta T_0$, somewhat underestimate the central values, $\Delta P_{\rm c}$ and $\Delta T_{\rm c}$, upon which Equation \ref{eqn:vortex_height} is based. Typically, though, encounters are only slightly offcenter \citep{2018Icar..299..166J}, and, in any case, we expect the uncertainties associated with the turbulent fluctuations in the temperature time series dominate. Inclusion of wind speed and direction measurements may help mitigate the miss distance effect, which are known to skew inferences about vortex properties \citep{2021PSJ.....2..206J}. 

The heights estimated here from meteorological data correspond to the altitude at which the rising convective plume radiates its excess heat to the environment, i.e., the atmospheric region representing the cold sink for the thermodynamic engine driving the convection. This height does not necessarily correspond to the top of a lofted dust column since dust may settle out of the flow below that height \citep{2011Icar..214..766M} or might have a low enough optical depth that seeing the top is difficult. These considerations suggest we might expect the heights estimated here to exceed heights estimated for dust devils visually. Instead, however, the visually measured dust devil heights seem to skew to larger values (the mode of heights reported in \citealt{2015Icar..260..246F} does agree, $500\,{\rm m}$, reasonably well with our median value, $450\,{\rm m}$). This disagreement seems likely to arise from a combination of limited image resolution and small number statistics, though. 

The discussion here shows that, as data from Perseverance continue pouring in, insights into Mars' PBL will come roaring in, too. The increased vortex encounter rates reported here promise growing scope for testing ideas about PBL processes, especially dust devils. As discussed here, a complete analysis of profiles of vortex pressures, temperatures, winds, and dust using the MEDA data could be transformative. Analysis of imagery from the mission also registers the passage of active dust devils \citep{2021DPS....5320307A}. Since the MEDA measurements represent only a single pass through a vortex, such a model could also substantially benefit from an terrestrial analog campaign involving a small network of weather stations to probe the same vortex through multiple lines of encounter \citep{2016Icar..271..326L, 2021DPS....5320304J}. 

\newpage

\movetabledown=60mm
\begin{rotatetable}
\begin{deluxetable}{cccccccccccccccc}
\tabletypesize{\scriptsize}
\tablecaption{Vortex parameters.\label{tbl:vortex_fit_params}}
\tablehead{\colhead{sol} & \colhead{$t_0^P$} & \colhead{$B_{\rm P}$} & \colhead{$S_{\rm P}$} & \colhead{$\Delta P_0$} & \colhead{${\rm SNR}_{\Delta P_0}$} & \colhead{$\Gamma_P$} & \colhead{ATS} & \colhead{$B_{\rm T}$} & \colhead{$S_{\rm P}$} & \colhead{$\Delta t_0^T$} & \colhead{$\Delta T_0$} & \colhead{${\rm SNR}_{\Delta T_0}$} & \colhead{$\Gamma_T$} & \colhead{Chain Converged?} & \colhead{$h/H$} \\ 
\colhead{} & \colhead{(LMST)} & \colhead{(Pa)} & \colhead{(${\rm Pa\ hr^{-1}}$)} & \colhead{(Pa)} & \colhead{} & \colhead{(s)} & \colhead{} & \colhead{(K)} & \colhead{(${\rm K\ hr^{-1}}$)} & \colhead{(s)} & \colhead{(K)} & \colhead{} & \colhead{(s)} & \colhead{} & \colhead{} } 
\startdata
15 &  $16:51:33 \pm 3 $ &  $723.22 \pm 0.04$ &  $-3 \pm 1$ &  $0.42 \pm 0.04$ &  $61 \pm 14$ &  9.3 &  3 &  $242.5 \pm 0.4$ &  $-0.001 \pm 0.002$ &  $2^{+10}_{-8}$ &  $1 \pm 1$ &  0.7 &  $74^{+63}_{-70}$ &  True &  - \\
15 &  $16:54:57.6 \pm 0.4 $ &  $723.14 \pm 0.05$ &  $7 \pm 6$ &  $0.98 \pm 0.07$ &  $12 \pm 2$ &  14.4 &  3 &  $242.2 \pm 0.1$ &  $0.008 \pm 0.004$ &  $1^{+0.7}_{-0.6}$ &  $2.5 \pm 0.5$ &  4.9 &  $8^{+2}_{-4}$ &  True &  $0.05 \pm 0.01$ \\
16 &  $13:19:32.8 \pm 0.7 $ &  $733.636 \pm 0.008$ &  $1.2 \pm 0.4$ &  $0.39 \pm 0.03$ &  $23 \pm 3$ &  15.1 &  3 &  $236 \pm 1$ &  $0.01 \pm 0.01$ &  $-0 \pm 2$ &  $9^{+1}_{-2}$ &  6.4 &  $59^{+11}_{-6}$ &  True &  - \\
\enddata
\tablecomments{Abbreviated version of vortex parameter table. The column labeled ``ATS'' shows which ATS sensor time series was used. That labeled ``Chain Converged'' shows whether the MCMC analysis of temperature time series converged.}
\end{deluxetable}
\end{rotatetable}

\newpage

\begin{acknowledgments}
This study benefited from conversations with Germ\'{a}n Mart\'{i}nez, Hartzel Gillespie, Susan Conway, Lori Fenton, Steve Metzger, Ryan Battin, Justin Crevier, and Hallie Dodge. This research was supported by a grant from NASA's Solar System Workings program, NNH17ZDA001N-SSW. This study also benefited from thoughtful and thorough feedback from two anonymous referees.
\end{acknowledgments}

\software{matplotlib \citep{Hunter:2007}, numpy \citep{harris2020array}, scipy \citep{2020SciPy-NMeth}, Astropy \citep{astropy:2013, astropy:2018}}

\appendix
\section{Corrected Derivation of Equation 1}
\label{sec:corrected_derivation}
In this appendix, we re-derive Equation \ref{eqn:vortex_height}. That equation is based on a derivation in \citet{2020Icar..33813523J}, but that latter derivation has a small error (which does not affect the results of that study). Relating the pressure excursion for a dust devil to the temperature excursion and other thermodynamic parameters, \citet{1998JAtS...55.3244R} derived the following equation
\begin{equation}
    \Delta P = P_{\infty} \left\{ 1 - \exp \left[ \left( \frac{\gamma \eta}{\gamma \eta - 1}\right) \left(\frac{1}{\chi}\right) \left( \frac{\Delta T}{T_{\infty}}\right) \right] \right\}\label{eqn:Renno_Delta-p}.
\end{equation}
Other variables are defined below Equation \ref{eqn:vortex_height}, but $\eta$ is the thermodynamic efficiency, given by 
\begin{equation}
    \eta = \frac{T_{\rm h} - T_{\rm c}}{T_{\rm h}}\label{eqn:Renno_eta},
\end{equation}{}
where $T_{\rm h}$ is the entropy-weighted mean temperature near the surface where heat is absorbed, and $T_{\rm c}$ is the same for the cold sink at the top of the dust devil. A useful approximation gives $T_{\rm h} \approx T_\infty$, while
\begin{equation}
    T_{\rm c} = \left[ \frac{ P_\infty^{\chi + 1} - P_{\rm top}^{\chi + 1} }{\left( P_\infty - P_{\rm top} \right) \left( \chi + 1\right) P_\infty^{\chi}} \right] T_{\rm h},\label{eqn:Tc}
\end{equation}
where $P_{\rm top}$ is the pressure near the top of the dust devil and is related to the surface pressure as $P_{\rm top} \approx P_\infty \exp\left(-h/H\right)$. \citet{2020Icar..33813523J} argued $h/H \ll 1$ and so expanded Equation \ref{eqn:Tc}, giving
\begin{equation}
    \eta \approx \left( \frac{1 + \chi}{2} \right) \left( \frac{h}{H} \right).\label{eqn:approx_eta}
\end{equation}
(In the original derivation in \citet{2020Icar..33813523J}, $1 + \chi$ was mistakenly written as $\chi$.) 

We can plug Equation \ref{eqn:approx_eta} into Equation \ref{eqn:Renno_Delta-p} and again expand about small $h/H$. Re-arranging the result to relate $h/H$ to the pressure and temperature excursions gives Equation \ref{eqn:vortex_height}.

\bibliography{sample631}{}

\begin{thebibliography}{}
\expandafter\ifx\csname natexlab\endcsname\relax\def\natexlab#1{#1}\fi
\providecommand{\url}[1]{\href{#1}{#1}}
\providecommand{\dodoi}[1]{doi:~\href{http://doi.org/#1}{\nolinkurl{#1}}}
\providecommand{\doeprint}[1]{\href{http://ascl.net/#1}{\nolinkurl{http://ascl.net/#1}}}
\providecommand{\doarXiv}[1]{\href{https://arxiv.org/abs/#1}{\nolinkurl{https://arxiv.org/abs/#1}}}

\bibitem[{{Allison} \& {McEwen}(2000)}]{2000P&SS...48..215A}
{Allison}, M., \& {McEwen}, M. 2000, \planss, 48, 215,
  \dodoi{10.1016/S0032-0633(99)00092-6}

\bibitem[{{Apestigue} {et~al.}(2021){Apestigue}, {Toledo}, {Arruego}, {Smith},
  {Lemmon}, {G{\'o}mez}, {Yela}, {Jim{\'e}nez}, {Garc{\'\i}a},
  {G{\'o}mez-Elvira}, {Navarro}, {Mart{\'\i}nez}, {Sebasti{\'a}n}, {de
  Vicente-Retortillo}, {S{\'a}nchez-Lavega}, {Hueso}, {Harri}, {Genzer},
  {Hieta}, {Polkko}, {Newman}, {De La Torre Juarez}, \&
  {Rodr{\'\i}guez-Manfredi}}]{2021DPS....5320307A}
{Apestigue}, V., {Toledo}, D., {Arruego}, I., {et~al.} 2021, in AAS/Division
  for Planetary Sciences Meeting Abstracts, Vol.~53, AAS/Division for Planetary
  Sciences Meeting Abstracts, 203.07

\bibitem[{{Arya}(2001)}]{2001imm..book.....A}
{Arya}, S.~P. 2001, {Introduction to Micrometeorology}

\bibitem[{{Astropy Collaboration} {et~al.}(2013){Astropy Collaboration},
  {Robitaille}, {Tollerud}, {Greenfield}, {Droettboom}, {Bray}, {Aldcroft},
  {Davis}, {Ginsburg}, {Price-Whelan}, {Kerzendorf}, {Conley}, {Crighton},
  {Barbary}, {Muna}, {Ferguson}, {Grollier}, {Parikh}, {Nair}, {Unther},
  {Deil}, {Woillez}, {Conseil}, {Kramer}, {Turner}, {Singer}, {Fox}, {Weaver},
  {Zabalza}, {Edwards}, {Azalee Bostroem}, {Burke}, {Casey}, {Crawford},
  {Dencheva}, {Ely}, {Jenness}, {Labrie}, {Lim}, {Pierfederici}, {Pontzen},
  {Ptak}, {Refsdal}, {Servillat}, \& {Streicher}}]{astropy:2013}
{Astropy Collaboration}, {Robitaille}, T.~P., {Tollerud}, E.~J., {et~al.} 2013,
  \aap, 558, A33, \dodoi{10.1051/0004-6361/201322068}

\bibitem[{{Astropy Collaboration} {et~al.}(2018){Astropy Collaboration},
  {Price-Whelan}, {Sip{\H{o}}cz}, {G{\"u}nther}, {Lim}, {Crawford}, {Conseil},
  {Shupe}, {Craig}, {Dencheva}, {Ginsburg}, {Vand erPlas}, {Bradley},
  {P{\'e}rez-Su{\'a}rez}, {de Val-Borro}, {Aldcroft}, {Cruz}, {Robitaille},
  {Tollerud}, {Ardelean}, {Babej}, {Bach}, {Bachetti}, {Bakanov}, {Bamford},
  {Barentsen}, {Barmby}, {Baumbach}, {Berry}, {Biscani}, {Boquien}, {Bostroem},
  {Bouma}, {Brammer}, {Bray}, {Breytenbach}, {Buddelmeijer}, {Burke},
  {Calderone}, {Cano Rodr{\'\i}guez}, {Cara}, {Cardoso}, {Cheedella}, {Copin},
  {Corrales}, {Crichton}, {D'Avella}, {Deil}, {Depagne}, {Dietrich}, {Donath},
  {Droettboom}, {Earl}, {Erben}, {Fabbro}, {Ferreira}, {Finethy}, {Fox},
  {Garrison}, {Gibbons}, {Goldstein}, {Gommers}, {Greco}, {Greenfield},
  {Groener}, {Grollier}, {Hagen}, {Hirst}, {Homeier}, {Horton}, {Hosseinzadeh},
  {Hu}, {Hunkeler}, {Ivezi{\'c}}, {Jain}, {Jenness}, {Kanarek}, {Kendrew},
  {Kern}, {Kerzendorf}, {Khvalko}, {King}, {Kirkby}, {Kulkarni}, {Kumar},
  {Lee}, {Lenz}, {Littlefair}, {Ma}, {Macleod}, {Mastropietro}, {McCully},
  {Montagnac}, {Morris}, {Mueller}, {Mumford}, {Muna}, {Murphy}, {Nelson},
  {Nguyen}, {Ninan}, {N{\"o}the}, {Ogaz}, {Oh}, {Parejko}, {Parley}, {Pascual},
  {Patil}, {Patil}, {Plunkett}, {Prochaska}, {Rastogi}, {Reddy Janga},
  {Sabater}, {Sakurikar}, {Seifert}, {Sherbert}, {Sherwood-Taylor}, {Shih},
  {Sick}, {Silbiger}, {Singanamalla}, {Singer}, {Sladen}, {Sooley},
  {Sornarajah}, {Streicher}, {Teuben}, {Thomas}, {Tremblay}, {Turner},
  {Terr{\'o}n}, {van Kerkwijk}, {de la Vega}, {Watkins}, {Weaver}, {Whitmore},
  {Woillez}, {Zabalza}, \& {Astropy Contributors}}]{astropy:2018}
{Astropy Collaboration}, {Price-Whelan}, A.~M., {Sip{\H{o}}cz}, B.~M., {et~al.}
  2018, \aj, 156, 123, \dodoi{10.3847/1538-3881/aabc4f}

\bibitem[{{Balme} \& {Greeley}(2006)}]{2006RvGeo..44.3003B}
{Balme}, M., \& {Greeley}, R. 2006, Reviews of Geophysics, 44, RG3003,
  \dodoi{10.1029/2005RG000188}

\bibitem[{{Cola{\"\i}tis} {et~al.}(2013){Cola{\"\i}tis}, {Spiga}, {Hourdin},
  {Rio}, {Forget}, \& {Millour}}]{2013JGRE..118.1468C}
{Cola{\"\i}tis}, A., {Spiga}, A., {Hourdin}, F., {et~al.} 2013, Journal of
  Geophysical Research (Planets), 118, 1468, \dodoi{10.1002/jgre.20104}

\bibitem[{{Ellehoj} {et~al.}(2010){Ellehoj}, {Gunnlaugsson}, {Taylor},
  {Kahanp{\"a}{\"a}}, {Bean}, {Cantor}, {Gheynani}, {Drube}, {Fisher}, {Harri},
  {Holstein-Rathlou}, {Lemmon}, {Madsen}, {Malin}, {Polkko}, {Smith},
  {Tamppari}, {Weng}, \& {Whiteway}}]{2010JGRE..115.0E16E}
{Ellehoj}, M.~D., {Gunnlaugsson}, H.~P., {Taylor}, P.~A., {et~al.} 2010,
  Journal of Geophysical Research (Planets), 115, E00E16,
  \dodoi{10.1029/2009JE003413}

\bibitem[{{Farley} {et~al.}(2020){Farley}, {Williford}, {Stack}, {Bhartia},
  {Chen}, {de la Torre}, {Hand}, {Goreva}, {Herd}, {Hueso}, {Liu}, {Maki},
  {Martinez}, {Moeller}, {Nelessen}, {Newman}, {Nunes}, {Ponce}, {Spanovich},
  {Willis}, {Beegle}, {Bell}, {Brown}, {Hamran}, {Hurowitz}, {Maurice},
  {Paige}, {Rodriguez-Manfredi}, {Schulte}, \& {Wiens}}]{2020SSRv..216..142F}
{Farley}, K.~A., {Williford}, K.~H., {Stack}, K.~M., {et~al.} 2020, \ssr, 216,
  142, \dodoi{10.1007/s11214-020-00762-y}

\bibitem[{{Fenton} {et~al.}(2016){Fenton}, {Reiss}, {Lemmon}, {Marticorena},
  {Lewis}, \& {Cantor}}]{2016SSRv..203...89F}
{Fenton}, L., {Reiss}, D., {Lemmon}, M., {et~al.} 2016, \ssr, 203, 89,
  \dodoi{10.1007/s11214-016-0243-6}

\bibitem[{{Fenton} \& {Lorenz}(2015)}]{2015Icar..260..246F}
{Fenton}, L.~K., \& {Lorenz}, R. 2015, \icarus, 260, 246,
  \dodoi{10.1016/j.icarus.2015.07.028}

\bibitem[{{Foreman-Mackey}(2018)}]{celerite2}
{Foreman-Mackey}, D. 2018, Research Notes of the American Astronomical Society,
  2, 31, \dodoi{10.3847/2515-5172/aaaf6c}

\bibitem[{{Foreman-Mackey} {et~al.}(2017){Foreman-Mackey}, {Agol},
  {Ambikasaran}, \& {Angus}}]{celerite1}
{Foreman-Mackey}, D., {Agol}, E., {Ambikasaran}, S., \& {Angus}, R. 2017, \aj,
  154, 220, \dodoi{10.3847/1538-3881/aa9332}

\bibitem[{{Foreman-Mackey} {et~al.}(2013){Foreman-Mackey}, {Hogg}, {Lang}, \&
  {Goodman}}]{2013PASP..125..306F}
{Foreman-Mackey}, D., {Hogg}, D.~W., {Lang}, D., \& {Goodman}, J. 2013, \pasp,
  125, 306, \dodoi{10.1086/670067}

\bibitem[{Harris {et~al.}(2020)Harris, Millman, van~der Walt, Gommers,
  Virtanen, Cournapeau, Wieser, Taylor, Berg, Smith, Kern, Picus, Hoyer, van
  Kerkwijk, Brett, Haldane, del R{'{\i}}o, Wiebe, Peterson,
  G{'{e}}rard-Marchant, Sheppard, Reddy, Weckesser, Abbasi, Gohlke, \&
  Oliphant}]{harris2020array}
Harris, C.~R., Millman, K.~J., van~der Walt, S.~J., {et~al.} 2020, Nature, 585,
  357, \dodoi{10.1038/s41586-020-2649-2}

\bibitem[{{Hess} \& {Spillane}(1990)}]{1990JApMe..29..498H}
{Hess}, G.~D., \& {Spillane}, K.~T. 1990, Journal of Applied Meteorology, 29,
  498

\bibitem[{Hunter(2007)}]{Hunter:2007}
Hunter, J.~D. 2007, Computing in Science \& Engineering, 9, 90,
  \dodoi{10.1109/MCSE.2007.55}

\bibitem[{{Jackson}(2020)}]{2020Icar..33813523J}
{Jackson}, B. 2020, \icarus, 338, 113523, \dodoi{10.1016/j.icarus.2019.113523}

\bibitem[{{Jackson}(2022)}]{2022PSJ.....3...20J}
---. 2022, \psj, 3, 20, \dodoi{10.3847/PSJ/ac4586}

\bibitem[{{Jackson} {et~al.}(2021{\natexlab{a}}){Jackson}, {Crevier},
  {Szurgot}, {Battin}, {Perrin}, \& {Rodriguez}}]{2021PSJ.....2..206J}
{Jackson}, B., {Crevier}, J., {Szurgot}, M., {et~al.} 2021{\natexlab{a}}, \psj,
  2, 206, \dodoi{10.3847/PSJ/ac260d}

\bibitem[{{Jackson} {et~al.}(2018{\natexlab{a}}){Jackson}, {Lorenz}, \&
  {Davis}}]{2018Icar..299..166J}
{Jackson}, B., {Lorenz}, R., \& {Davis}, K. 2018{\natexlab{a}}, \icarus, 299,
  166, \dodoi{10.1016/j.icarus.2017.07.027}

\bibitem[{{Jackson} {et~al.}(2018{\natexlab{b}}){Jackson}, {Lorenz}, {Davis},
  \& {Lipple}}]{2018RemS...10...65J}
{Jackson}, B., {Lorenz}, R., {Davis}, K., \& {Lipple}, B. 2018{\natexlab{b}},
  Remote Sensing, 10, 65, \dodoi{10.3390/rs10010065}

\bibitem[{{Jackson} {et~al.}(2021{\natexlab{b}}){Jackson}, {Lorenz}, {Fenton},
  {Crevier}, {Battin}, \& {Szurgot}}]{2021DPS....5320304J}
{Jackson}, B., {Lorenz}, R., {Fenton}, L., {et~al.} 2021{\natexlab{b}}, in
  AAS/Division for Planetary Sciences Meeting Abstracts, Vol.~53, AAS/Division
  for Planetary Sciences Meeting Abstracts, 203.04

\bibitem[{{Kurgansky}(2020)}]{2020Icar..33513389K}
{Kurgansky}, M.~V. 2020, \icarus, 335, 113389,
  \dodoi{10.1016/j.icarus.2019.113389}

\bibitem[{{Kurgansky}(2021)}]{2021Icar..35814200K}
---. 2021, \icarus, 358, 114200, \dodoi{10.1016/j.icarus.2020.114200}

\bibitem[{{Lorenz}(2011)}]{2011Icar..215..381L}
{Lorenz}, R. 2011, \icarus, 215, 381, \dodoi{10.1016/j.icarus.2011.06.005}

\bibitem[{{Lorenz}(2014)}]{2014JAtS...71.4461L}
{Lorenz}, R.~D. 2014, Journal of Atmospheric Sciences, 71, 4461,
  \dodoi{10.1175/JAS-D-14-0138.1}

\bibitem[{{Lorenz}(2016)}]{2016Icar..271..326L}
---. 2016, \icarus, 271, 326, \dodoi{10.1016/j.icarus.2016.02.001}

\bibitem[{{Lorenz}(2021)}]{2021Icar..35414062L}
---. 2021, \icarus, 354, 114062, \dodoi{10.1016/j.icarus.2020.114062}

\bibitem[{{Lorenz} {et~al.}(2021){Lorenz}, {Spiga}, {Lognonn{\'e}}, {Plasman},
  {Newman}, \& {Charalambous}}]{2021Icar..35514119L}
{Lorenz}, R.~D., {Spiga}, A., {Lognonn{\'e}}, P., {et~al.} 2021, \icarus, 355,
  114119, \dodoi{10.1016/j.icarus.2020.114119}

\bibitem[{{Metzger} {et~al.}(2011){Metzger}, {Balme}, {Towner}, {Bos},
  {Ringrose}, \& {Patel}}]{2011Icar..214..766M}
{Metzger}, S.~M., {Balme}, M.~R., {Towner}, M.~C., {et~al.} 2011, \icarus, 214,
  766, \dodoi{10.1016/j.icarus.2011.03.013}

\bibitem[{{Murphy} {et~al.}(2016){Murphy}, {Steakley}, {Balme}, {Deprez},
  {Esposito}, {Kahanp{\"a}{\"a}}, {Lemmon}, {Lorenz}, {Murdoch}, {Neakrase},
  {Patel}, \& {Whelley}}]{2016SSRv..203...39M}
{Murphy}, J., {Steakley}, K., {Balme}, M., {et~al.} 2016, \ssr, 203, 39,
  \dodoi{10.1007/s11214-016-0283-y}

\bibitem[{{Newman} {et~al.}(2021){Newman}, {de la Torre Ju{\'a}rez},
  {Pla-Garc{\'\i}a}, {Wilson}, {Lewis}, {Neary}, {Kahre}, {Forget}, {Spiga},
  {Richardson}, {Daerden}, {Bertrand}, {Vi{\'u}dez-Moreiras}, {Sullivan},
  {S{\'a}nchez-Lavega}, {Chide}, \& {Rodriguez-Manfredi}}]{2021SSRv..217...20N}
{Newman}, C.~E., {de la Torre Ju{\'a}rez}, M., {Pla-Garc{\'\i}a}, J., {et~al.}
  2021, \ssr, 217, 20, \dodoi{10.1007/s11214-020-00788-2}

\bibitem[{Newman {et~al.}(2022)Newman, Hueso, Lemmon, Munguira, Álvaro
  Vicente-Retortillo, Apestigue, Martínez, Toledo, Sullivan, Herkenhoff, de~la
  Torre~Juárez, Richardson, Stott, Murdoch, Sanchez-Lavega, Wolff, Arruego,
  Sebastián, Navarro, Gómez-Elvira, Tamppari, Viúdez-Moreiras, Harri,
  Genzer, Hieta, Lorenz, Conrad, Gómez, McConnochie, Mimoun, Tate, Bertrand,
  Bell, Maki, Rodriguez-Manfredi, Wiens, Chide, Maurice, Zorzano, Mora, Baker,
  Banfield, Pla-Garcia, Beyssac, Brown, Clark, Lepinette, Montmessin, Fischer,
  Patel, del Río-Gaztelurrutia, Fouchet, Francis, \& Guzewich}]{Newman2022}
Newman, C.~E., Hueso, R., Lemmon, M.~T., {et~al.} 2022, Science Advances, 8,
  eabn3783, \dodoi{10.1126/sciadv.abn3783}

\bibitem[{{Petrosyan} {et~al.}(2011){Petrosyan}, {Galperin}, {Larsen}, {Lewis},
  {M{\"a}{\"a}tt{\"a}nen}, {Read}, {Renno}, {Rogberg}, {Savij{\"a}rvi},
  {Siili}, {Spiga}, {Toigo}, \& {V{\'a}zquez}}]{2011RvGeo..49.3005P}
{Petrosyan}, A., {Galperin}, B., {Larsen}, S.~E., {et~al.} 2011, Reviews of
  Geophysics, 49, RG3005, \dodoi{10.1029/2010RG000351}

\bibitem[{{Pla-Garc{\'\i}a} {et~al.}(2020){Pla-Garc{\'\i}a}, {Rafkin},
  {Martinez}, {Vicente-Retortillo}, {Newman}, {Savij{\"a}rvi}, {de la Torre},
  {Rodriguez-Manfredi}, {G{\'o}mez}, {Molina}, {Vi{\'u}dez-Moreiras}, \&
  {Harri}}]{2020SSRv..216..148P}
{Pla-Garc{\'\i}a}, J., {Rafkin}, S.~C.~R., {Martinez}, G.~M., {et~al.} 2020,
  \ssr, 216, 148, \dodoi{10.1007/s11214-020-00763-x}

\bibitem[{Press {et~al.}(2007)Press, Teukolsky, Vetterling, \&
  Flannery}]{Press2007}
Press, W.~H., Teukolsky, S.~A., Vetterling, W.~T., \& Flannery, B.~P. 2007,
  Numerical Recipes 3rd Edition: The Art of Scientific Computing, 3rd edn.
  (Cambridge University Press).
\newblock
  \url{http://www.amazon.com/Numerical-Recipes-3rd-Scientific-Computing/dp/0521880688/ref=sr_1_1?ie=UTF8&s=books&qid=1280322496&sr=8-1}

\bibitem[{Rasmussen \& Williams(2006)}]{books/lib/RasmussenW06}
Rasmussen, C.~E., \& Williams, C. K.~I. 2006, Gaussian processes for machine
  learning., Adaptive computation and machine learning (MIT Press), I--XVIII,
  1--248

\bibitem[{{Reiss} {et~al.}(2014){Reiss}, {Spiga}, \&
  {Erkeling}}]{2014Icar..227....8R}
{Reiss}, D., {Spiga}, A., \& {Erkeling}, G. 2014, \icarus, 227, 8,
  \dodoi{10.1016/j.icarus.2013.08.028}

\bibitem[{{Renn{\'o}} {et~al.}(1998){Renn{\'o}}, {Burkett}, \&
  {Larkin}}]{1998JAtS...55.3244R}
{Renn{\'o}}, N.~O., {Burkett}, M.~L., \& {Larkin}, M.~P. 1998, Journal of
  Atmospheric Sciences, 55, 3244,
  \dodoi{10.1175/1520-0469(1998)055<3244:ASTTFD>2.0.CO;2}

\bibitem[{{Renn{\'o}} {et~al.}(2000){Renn{\'o}}, {Nash}, {Lunine}, \&
  {Murphy}}]{2000JGR...105.1859R}
{Renn{\'o}}, N.~O., {Nash}, A.~A., {Lunine}, J., \& {Murphy}, J. 2000, \jgr,
  105, 1859, \dodoi{10.1029/1999JE001037}

\bibitem[{{Ringrose} {et~al.}(2003){Ringrose}, {Towner}, \&
  {Zarnecki}}]{2003Icar..163...78R}
{Ringrose}, T.~J., {Towner}, M.~C., \& {Zarnecki}, J.~C. 2003, \icarus, 163,
  78, \dodoi{10.1016/S0019-1035(03)00073-3}

\bibitem[{{Rodriguez-Manfredi} {et~al.}(2021){Rodriguez-Manfredi}, {de la Torre
  Ju{\'a}rez}, {Alonso}, {Ap{\'e}stigue}, {Arruego}, {Atienza}, {Banfield},
  {Boland}, {Carrera}, {Casta{\~n}er}, {Ceballos}, {Chen-Chen}, {Cobos},
  {Conrad}, {Cordoba}, {del R{\'\i}o-Gaztelurrutia}, {de Vicente-Retortillo},
  {Dom{\'\i}nguez-Pumar}, {Espejo}, {Fairen}, {Fern{\'a}ndez-Palma},
  {Ferr{\'a}ndiz}, {Ferri}, {Fischer}, {Garc{\'\i}a-Manchado},
  {Garc{\'\i}a-Villadangos}, {Genzer}, {Gim{\'e}nez}, {G{\'o}mez-Elvira},
  {G{\'o}mez}, {Guzewich}, {Harri}, {Hern{\'a}ndez}, {Hieta}, {Hueso},
  {Jaakonaho}, {Jim{\'e}nez}, {Jim{\'e}nez}, {Larman}, {Leiter}, {Lepinette},
  {Lemmon}, {L{\'o}pez}, {Madsen}, {M{\"a}kinen}, {Mar{\'\i}n},
  {Mart{\'\i}n-Soler}, {Mart{\'\i}nez}, {Molina}, {Mora-Sotomayor},
  {Moreno-{\'A}lvarez}, {Navarro}, {Newman}, {Ortega}, {Parrondo}, {Peinado},
  {Pe{\~n}a}, {P{\'e}rez-Grande}, {P{\'e}rez-Hoyos}, {Pla-Garc{\'\i}a},
  {Polkko}, {Postigo}, {Prieto-Ballesteros}, {Rafkin}, {Ramos}, {Richardson},
  {Romeral}, {Romero}, {Runyon}, {Saiz-Lopez}, {S{\'a}nchez-Lavega}, {Sard},
  {Schofield}, {Sebastian}, {Smith}, {Sullivan}, {Tamppari}, {Thompson},
  {Toledo}, {Torrero}, {Torres}, {Urqu{\'\i}}, {Velasco},
  {Vi{\'u}dez-Moreiras}, {Zurita}, \& {MEDA team}}]{2021SSRv..217...48R}
{Rodriguez-Manfredi}, J.~A., {de la Torre Ju{\'a}rez}, M., {Alonso}, A.,
  {et~al.} 2021, \ssr, 217, 48, \dodoi{10.1007/s11214-021-00816-9}

\bibitem[{{Ryan} \& {Carroll}(1970)}]{1970JGR....75..531R}
{Ryan}, J.~A., \& {Carroll}, J.~J. 1970, \jgr, 75, 531,
  \dodoi{10.1029/JC075i003p00531}

\bibitem[{{Spiga} {et~al.}(2021){Spiga}, {Murdoch}, {Lorenz}, {Forget},
  {Newman}, {Rodriguez}, {Pla-Garcia}, {Moreiras}, {Banfield}, {Perrin},
  {Mueller}, {Lemmon}, {Millour}, \& {Banerdt}}]{Spiga2021}
{Spiga}, A., {Murdoch}, N., {Lorenz}, R., {et~al.} 2021, Journal of Geophysical
  Research (Planets), 126, e06511, \dodoi{10.1029/2020JE006511}

\bibitem[{{Stanzel} {et~al.}(2008){Stanzel}, {P{\"a}tzold}, {Williams},
  {Whelley}, {Greeley}, {Neukum}, \& {HRSC Co-Investigator
  Team}}]{2008Icar..197...39S}
{Stanzel}, C., {P{\"a}tzold}, M., {Williams}, D.~A., {et~al.} 2008, \icarus,
  197, 39, \dodoi{10.1016/j.icarus.2008.04.017}

\bibitem[{{Steakley} \& {Murphy}(2016)}]{2016Icar..278..180S}
{Steakley}, K., \& {Murphy}, J. 2016, \icarus, 278, 180,
  \dodoi{10.1016/j.icarus.2016.06.010}

\bibitem[{Virtanen {et~al.}(2020)Virtanen, Gommers, Oliphant, Haberland, Reddy,
  Cournapeau, Burovski, Peterson, Weckesser, Bright, {van der Walt}, Brett,
  Wilson, Millman, Mayorov, Nelson, Jones, Kern, Larson, Carey, Polat, Feng,
  Moore, {VanderPlas}, Laxalde, Perktold, Cimrman, Henriksen, Quintero, Harris,
  Archibald, Ribeiro, Pedregosa, {van Mulbregt}, \& {SciPy 1.0
  Contributors}}]{2020SciPy-NMeth}
Virtanen, P., Gommers, R., Oliphant, T.~E., {et~al.} 2020, Nature Methods, 17,
  261, \dodoi{10.1038/s41592-019-0686-2}

\end{thebibliography}
\bibliographystyle{aasjournal}

\end{document}